\documentclass[12pt,reqno]{amsart}
\usepackage{amsmath,amsxtra,amssymb,amsthm,amsfonts,eufrak}

\usepackage{tikz-cd}
\makeatletter
\newcommand*{\rom}[1]{\expandafter\@slowromancap\romannumeral #1@}
\makeatother
\usetikzlibrary{matrix,arrows,decorations.pathmorphing}

\numberwithin{equation}{section}

\newcommand{\gl}{\pl \ge \pl}

\newcommand{\lel}{\pl = \pl}

\newcommand{\R}{{\mathcal R}}

\newcommand{\ten}{\otimes}

\newcommand{\pl}{\hspace{.1cm}}

\newcommand{\ran}{\rangle}

\newcommand{\al}{\alpha}
\newcommand{\si}{\sigma}

\newcommand{\la}{\lambda}
\newcommand{\eps}{\varepsilon}

\newcommand{\id}{\iota_{\infty,2}^n}

\newcommand{\A}{{\mathcal A}}
\newcommand{\B}{{\mathcal{B}}}
\newcommand{\bB}{{\mathbb{B}}}

\newcommand{\M}{{\mathcal M}}

\newcommand{\N}{{\mathcal N}}

\newcommand{\norm}[2]{\parallel \! #1 \! \parallel_{#2}}

\newtheorem{lemma}{Lemma}[section]

\newtheorem{theorem}[lemma]{Theorem}
\newtheorem{cor}[lemma]{Corollary}

\newtheorem{rem}[lemma]{Remark}

\newcommand{\re}{\begin{rem}\rm}
\newcommand{\mar}{\end{rem}}
\newtheorem{exam}[lemma]{Example}
\newcommand{\bra}[1]{\langle{#1}|}
\newcommand{\ket}[1]{|{#1}\rangle}
\newcommand{\ketbra}[1]{|{#1}\rangle\langle{#1}|}

\newcommand{\qd}{\end{proof}\vspace{0.5ex}}
\newcommand{\prf}{\begin{proof}[\bf Proof:]}

\newcommand{\xspace}{\hbox{\kern-2.5pt}}

\newcommand{\tr}{{{\text{tr}}}}
\renewcommand{\id}{{\text{id}}}

\newcommand{\cA}{{\mathcal A}}
\newcommand{\cB}{{\mathcal B}}

\newcommand{\cD}{{\mathcal D}}

\newcommand{\cM}{{\mathcal M}}

\newcommand{\cR}{{\mathcal R}}

\renewcommand{\bB}{{\mathbb B}}

\newcommand{\bM}{{\mathbb M}}

\newcommand{\bR}{{\mathbb R}}

\newtheorem*{theorem*}{Theorem}

\allowdisplaybreaks

\begin{document}
\title{Entropy Uncertainty Relations and Strong sub-additivity of Quantum Channels}
\author{Li Gao}
\address{Department of Mathematics\\
University of Houston} \email[Li Gao]{lgao12@uh.edu}
\author{Marius Junge}
\address{Department of Mathematics\\
University of Illinois, Urbana, IL 61801, USA} \email[Marius Junge]{mjunge@illinois.edu}
\author{Nicholas LaRacuente}
\address{Chicago Quantum Exchange, University of Chicago, Chicago, IL 60637, USA} \email[Nicholas LaRacuente]{nlaracuente@uchicago.edu}
\maketitle

\begin{abstract}
We prove an entropic uncertainty relation for two quantum channels, extending the work of Frank and Lieb for quantum measurements. This is obtained via a generalized strong super-additivity (SSA) of quantum entropy. Motivated by Petz's algebraic SSA inequality, we also obtain a generalized SSA for quantum relative entropy. As a special case, it gives an improved data processing inequality.
\end{abstract}

\section{Introduction}
Uncertainty principle is a fundamental phenomenon in quantum mechanics. The celebrated
Heisenberg's uncertainty principle states that the position and momentum of a quantum particle cannot be sharply defined at same time, i.e.
\begin{align}\label{eq:heisenberg}\sigma(Q)\sigma(P)\ge \frac{\hbar}{2}\pl,\end{align}
where $\sigma(Q)$ and $\sigma(P)$ denote the standard derivation of the position and momentum respectively, and $\hbar$ is the reduced Planck constant \cite{kennard1927quantenmechanik,weyl1928gruppentheorie}. Such uncertainty relations widely exist in quantum physics, such as energy-. In fact, for two observables described by Hermitian operators $X$ and $Z$, Robertson \cite{robertson1929uncertainty} proved that
\begin{align} \sigma(X)\sigma(Z)\ge \frac{1}{2}| \bra{\psi} [X,Z]  \ket{\psi} |\pl, \label{eq:robertson}\end{align}
where $\ket{\psi}$ is the state of the quantum system and $[\cdot ,\cdot ]$ denotes the commutator. The Heisenberg's principle \eqref{eq:heisenberg} is then a consequence for the commutation relation $[Q,P]=-i\hbar I$. Robertson's inequality shows that  uncertainty principle \eqref{eq:robertson} is a reflection of non-commutativity, which is an essential feature of quantum physics.

In statistical physics and information theory, entropy is a natural measure of uncertainty. Since Hirschman's first work \cite{hirschman1957note} on entropic uncertainty relation, there has been a series of works on uncertainty principle via entropic quantities (see the survey \cite{coles2017entropic} and the references therein). A notable one is that for the position $Q$ and momentum $P$,
\begin{align} h(Q)+h(P)\ge \log (e\pi \hbar)\pl,\label{eq:beckner}\end{align}
where $h(Q)=\int_{\mathbb{R}}\frac{dQ}{dq}\log \frac{dQ}{dq}dq$ is the differential entropy and $\frac{dQ}{dq}$ is the probability density function w.r.t the Lesbegue measure $dq$. This inequality was proved by Beckner \cite{beckner1975inequalities}, and also by Bialynicki-Birula and Mycielski \cite{bialynicki1975uncertainty} using sharp Hausdorff-Young inequality \cite{beckner1975inequalities}. Moreover, it is known to be stronger than Heisenberg's principle \eqref{eq:heisenberg} of standard deviation. For two observables $X$ and $Z$ of finite spectrum,
 Maassen and Uffink \cite{maassen1988generalized} discovered that
 \begin{align} H(X)+H(Z)\ge \log\frac{1}{c}\pl,\label{eq:MU}\end{align}
 where $H(X)=-\sum_{x} P_X(x)\log P_X(x)$ is the Shannon entropy, and $c=\max_{x,z}|\bra{x}z\ran|^2$ is the maximum overlap between the eigenbasis $\{\ket{x}\}_{x\in \mathcal{X}}$ and $\{\ket{z}\}_{z\in\mathcal{Z}}$ of $X$ and $Z$ respectively. A recent breakthrough was made by Berta \emph{et al}, which extends Maassen-Uffink relation to mixed quantum states, and more importantly, in the presence of quantum memory \cite{berta2010uncertainty}.

 Let $H_M$ be as Hilbert space and $\bB(H_M)$ be the bounded operator on $H_M$. A mixed quantum state on $H_M$ is modelled by a positive and trace 1 element $\rho$, called a density operator. Its von Neumann entropy is defined as \[H(\rho)=-\tr(\rho\log\rho)\pl,\] where $\tr$ is matrix trace. Berta \emph{et al} showed that any joint quantum state $\rho_{MC}$ on $H_M\ten H_C$ satisfies the following uncertainty relation,
\begin{align}\label{eq:berta} H(X|C)+H(Z|C)\ge H(M|C)+ \log\frac{1}{c}\pl.\end{align}
Here $C$ is a quantum reference system, $H(M|C)=H(\rho_{MC})-H(\rho_{C})$ is the conditional entropy of $\rho_{MC}$ with respect to the system $C$ (similarly, for $H(X|C)$ and $H(Z|C)$). It is worth noting that the constant $c$ is independent of the system $C$. This result has been further generalized to arbitrary measurements by Frank and Lieb \cite{frank2013extended}. Recall that a positive operator value measurement (POVM) on $H_M$ is a family of positive operators $\{E_x\}$ such that $\sum_x E_x=I$. Frank and Lieb  \cite{frank2013extended}  proved that: given two POVMs $\{E_x\}$ and $\{F_z\}$, any joint quantum state $\rho_{MC}$ satisfy
\eqref{eq:MU} with constant $c=\max_{x,z}\tr(E_xF_z)$, called the maximum overlap of measurements.

In this paper, we consider the entropy uncertainty relation for two quantum channels. Mathematically, a quantum  channel is a completely positive trace preserving map, which sends density operators to density operators. For simplicity, we only consider finite dimensional cases.

\begin{theorem*}[A]\label{thm:main1}
Let $H_A$, $H_B$ and $H_M$ be finite dimensional Hilbert spaces and $\Phi_A:\bB(H_M)\to \bB(H_A)\pl, \Phi_B:\bB(H_M)\to \bB(H_B)$ be two quantum channels. Then for any Hilbert space $H_C$ and any bipartite quantum state $\rho\in \bB(H_M\ten H_C)$
\[ H(A|C)_{\Phi_A(\rho)}+H(B|C)_{\Phi_B(\rho)}\ge H(M|C)_\rho+ \log\frac{1}{c} \pl.\]
The constant $c$ is given by the completely bounded norm
\begin{align} c=\norm{\Phi_B\circ \Phi_A^\dagger:S_1(H_A)\to \bB(H_B) }{cb}\pl,\end{align}
where $S_1(H_A)$ is the trace class operator on $H_A$ and $\Phi_A^\dagger$ is the adjoint map of $\Phi_A$.
 \end{theorem*}

  Note that by Effros-Ruan's isomorphism \cite{effros1990approximation,blecher1991tensor}, the constant $c$ equals to the operator norm of Choi matrix of $\Phi_B\circ \Phi_A^\dagger$, which is always finite. When the range of $\Phi_A$ and $\Phi_B$ are classical (commutative) systems, Theorem \ref{thm:main1} recovers the Frank-Lieb uncertainty relation. From this perspective, Theorem \ref{thm:main1} is a noncommutative generalization of Frank-Lieb's relation by allowing $\Phi_A$ and $\Phi_B$ to be quantum to quantum channels. Another special case is when $H_M=H_A\ten H_B$ and $\Phi_A=\id_A\ten \tr_B,\Phi_B=\tr_A\ten \id$ being the partial traces: this recovers the strong sub-additivity (SSA) of von Neumann entropy proved by  Lieb and Ruskai \cite{lieb1973proof},
 \[ H(AC)+H(BC)-H(ABC)-H(C)\ge 0\pl.\]
 In fact, our Theorem A is derived from the following generalized SSA inequality.

\begin{theorem*}[B]
\label{thm:main2}Let $\cA,\cB,\cM$ and $\cR$ be finite dimensional von Neumann algebras equipped with trace $\tau_\A,\tau_\B,\tau_\M$ and $\tau_\R$ respectively. Suppose $\cR\subset \cA$ as a subalgebra, and denote $E_R$ as the adjoint of the inclusion map. Given two quantum channels $\Phi_A:\M\to \A$ and $\Phi_B:\M\to \B$, for any density operator $\rho\in\cM$,
\begin{align}H(\Phi_A(\rho))+H(\Phi_B(\rho))\ge H(\rho)+H(E_\R\circ \Phi_A(\rho))+\log \frac{1}{c}\pl,\end{align}
where the constant $c$ is given by
\[c=\sup \{\tau_\M(\Phi_A^\dag(a)\Phi_B^\dag(b)) \pl |\pl a\in \A_+, b\in \B_+ \pl \pl,  E_\R(a)= 1\pl, \tau_\B(b)=1 \}\pl,\]
\end{theorem*}

Theorem B extends the algebraic SSA of Petz \cite{petz1991certain}: when $\cR\subset\cA,\cB\subset\M$ are subalgebras, $\Phi_A=E_A,\Phi_B=E_B$ are trace preserving conditional expectation, if $E_A\circ E_B=E_B\circ E_A=E_R $,
then
\[ H(E_A(\rho))+H(E_B(\rho))\ge H(\rho)+H(E_R\circ E_A(\rho))\pl.\]
The condition $E_\cA\circ E_\cB=E_\cB\circ E_\cA=E_\cR$ , called a commuting square, was first introduced by Popa \cite{popa1983orthogonal} , which is an important tool in the study of subfactors. Here, our constant $c=1$ if and only if the commuting square holds. From this perspective, Theorem B gives an entropic characterization for commuting square.

Motivated by Petz's algebraic SSA \cite[Theorem 12]{petz1991certain}, our third result is a generalized SSA for relative entropy. Recall that for two density operators $\rho$ and $\sigma$, the relative entropy is defined as $D(\rho||\sigma):=\tr(\rho\log\rho-\rho\log\sigma)$.

\begin{theorem*}[C]\label{thm:main3}Let $\Phi_A: \M\to \A, \Phi_B: \M\to \B$ be two quantum channels and $\R \subset\B$ is a subalgebra. Assume that $\sigma\in\M$ is a density operator and there exists a conditional expectation $E_R^\dagger:\B\to\R$ preserving the state $\Phi_B(\sigma)$. Then for any quantum state $\rho\in \M$, we have
\[D(\rho||\si)+D(E_R\circ\Phi_B(\rho)||E_R\circ\Phi_B(\si))\ge D(\Phi_A(\rho)||\Phi_A(\si))+D(\Phi_B(\rho)||\Phi_B(\si))-\kappa\]
The constant  $\kappa$  is given by
\begin{align*}
&\kappa=\int_{\bR}\al(t)\log c(t)dt\pl , \al(t)=\frac{\pi}{2(\cosh(\pi t) +1)}  \\
&c(t)=\sup_{b} \tau_\M\Big(\Phi_B^\dag(b)
\Phi_\A^\dag\big(\Phi_\A(\rho)^{\frac{1+it}{2}}\Phi_A(\si)^{\frac{-1-it}{2}}
\big)\si\Phi_\A^\dag\big(\Phi_\A(\rho)^{\frac{1+it}{2}}\Phi_A(\si)^{\frac{-1-it}{2}}
\big)^*\Big)
\end{align*}
where the supremum is for all $b\in \B_+$  such that  $E_R^\dagger(b)=1$.
\end{theorem*}
In particular, the above theorem gives an improvement of data processing inequality when $\A=\mathbb{C}$ and $\R=\mathbb{C}$ are trivial system.

The rest of paper is organized as follows. In Section 2, we discuss the connection between entropic quantities and noncommutative $L_p$-norms. Section 3 use complex interpolation of $L_p$-spaces to prove Theorem B, which differs with method of  Frank and Lieb for uncertainty relation of measurements. Section 4 reviews the operator space structure of noncommutative $L_p$-spaces and derive Theorem A. Section 5 discusses Petz's relative entropy SSA and prove Theorem C.\\

\noindent {\bf Notations}: We use italic letters $\cA,\cB,\cM,\cR\cdots $ for von Neumann algebras and subscript letter to index Hilbert space $H_A,H_B,H_C \cdots$. We will often use the short notation $H_{AB}=H_A\ten H_B$ for the tensor product space.
Given a finite dimensional Hilbert space $H$,
we denote $\mathbb{B}(H)$ (resp. $\mathbb{B}(H)_+$) as the set of bounded operators (resp. positive operators), and $\tr$ as the standard matrix trace. We use $1$ for the identity operator in $\bB(H)$ and $\id$ for the identity map. We write $A^*$ as the adjoint of an operator $A$ and $\Phi^\dagger$ as the adjoint of a map $\Phi$ with respect to trace inner product. \\

\noindent {\bf Acknowledgement}: LG is partially supported by NSF grant DMS-2154903. NL is supported as an IBM Postdoc at The University of Chicago. MJ was partially supported by NSF Grant DMS 1800872 and NSF RAISE-TAQS 1839177.\\

\noindent{\bf Note}: Theorem A is announced in the conference proceeding \cite{gao2018uncertainty} of \emph{IEEE International Symposium on Information Theory 2018}. Theorem B, Theorem C, as well as the proof of Theorem A in this paper are new.

\section{Entropy and $L_p$-norm}
\subsection{Noncommutative $L_p$-norm and von Neumann entropy}
We briefly review the connection between entropies and $L_p$-norms. The readers are referred to the survey \cite{pisier2003non} for more information on noncommutative $L_p$-space. For simplicity, throughout the paper we restrict ourselves to finite dimensional von Neumann algebras, i.e. $*$-subalgebras of matrix algebras. Let $\M$ be a finite dimensional von Neumann algebra and $\tau$ be a faithful trace on $\M$. For $0< p< \infty$, the non-commutative $L_p$-norm is defined
\[\norm{a}{L_p(\M,\tau)}=\tau(|a|^p)^{1/p} \pl, \pl a\in \M\pl,\]
and we denote by $L_p(\M,\tau)$ or simply $L_p(\M)$ for the $L_p$-space. In particular, $L_\infty(\M):=\M$. The basic example is Schatten $p$-class $S_p(H)=L_p(\bB(H),\tr)$, which is the $L_p$-space of $\bB(H)$ with respect to the matrix trace $\tr$. As classical $L_p$-spaces, non-commutative $L_p$-spaces forms a complex interpolation family,
\[ L_{p_\theta}(\M)=[L_{p_0}( \M),L_{p_1}( \M)]_\theta\pl,\]
where $ \frac{1}{p_\theta}=\frac{1-\theta}{p_0}+\frac{\theta}{p_1}$ and  $1\le p_0\le p_1\le \infty$. We refer to \cite{bergh2012interpolation} for the definition of complex interpolation.

The (quantum) states on $\M$ are given by density operators, which are positive and trace $1$. We denote
\[D(\M)=\{\rho\in \M \pl |\pl \rho\ge 0\pl, \pl \tau(\rho)=1 \}\pl,\pl D_+(\M)= \{ \rho\in D(\M)| \rho>0\}\]
as the state space and faithful state space respectively. The von Neumann entropy of a quantum state $\rho$ is defined as
\[H(\rho)=-\tau(\rho \log \rho) \pl. \]
This definition naturally extends to all positive operators. In general, $H(\rho)$ can be either negative or positive. Indeed, if the trace differs by a constant factor,
\[ \tilde{\tau}_\M=\lambda \tau_\M\pl, \pl \tilde{\rho} =\lambda^{-1}\rho  \pl,\]
the von Neumann entropy is up to a global constant
\[ H_\tau(\rho)=H_{\tilde{\tau}}(\tilde{\rho})+\log\lambda\pl.\]
\begin{exam}{\rm i) For the matrix trace $(\bB(H),\tr)$, $H(\rho)\ge 0$.\\
ii) For normalized trace $\tau(1)=1$, $H(\rho)\le 0$. \\
iii) Consider $L_\infty(\bR, dx)$ equipped with Lesbegue measure, $h(f)=-\int_{\bR} f(x)\log f(x) dx$ is called differential entropy, which can be either positive or negative. }
\end{exam}
The connection between von Neumann entropy and $L_p$-norm is as follows:
\begin{lemma}\label{lemma:1} i) For $\rho\in \M_+$,
\begin{align*}
&\lim_{p\to 1}\frac{\tau(\rho^p)-\tau(\rho)}{p-1}=-H(\rho)\\
&\lim_{p\to 1}\frac{\norm{\rho}{p}-\norm{\rho}{1}}{p-1}=-H(\rho)-\tau(\rho)\log \tau(\rho)\pl,
\end{align*}
and the two limits converges uniformly on $D(\M)$.\\
ii) If the path $\rho:[1,1+\eps)\to D(\M)$ satisfies $\displaystyle \lim_{p\to 1^+}\rho(p)=\rho$, then
\begin{align*}
\lim_{p\to 1^+}\frac{\tau(\rho(p)^p)-1}{p-1}=\lim_{p\to 1^+}\frac{\norm{\rho(p)}{p}-1}{p-1}=-H(\rho)
\end{align*}
\end{lemma}
\begin{proof}
For the first limit, we note that for positive number $x>0$, $p\mapsto \frac{x^p-1}{p-1}$ is monotone increasing and $\displaystyle \lim_{p\to 1}\frac{x^p-1}{p-1}=x\log x$. By monotone convergence theorem, \[ \lim_{p\to 1^+}\frac{\tau(\rho^p)-\tau(\rho)}{p-1}=\lim_{p\to 1^+}\frac{\tau(\rho^p-\rho)}{p-1}=\tau(\rho\log \rho)\pl. \]
In finite dimensions, $D(\M)$ is a compact set, hence by Dini's theorem, the convergence on $D(\M)$ is uniform. For the second limit, define the function
\[f(p)=\tau(\rho^p)\pl, p\in [1,\infty)\]
Then $f$ is continuously differentiable, $f(1)=\tau(\rho)$ and $f'(1^+)=-H(\rho)$. Using L'H\^{o}pital rule,
\begin{align*} &\lim_{p\to 1^+}\frac{f(p)^{\frac{1}{p}}-f(1)}{p-1}\\ = &f(1) \Big(-\log f(1)+\frac{f'(1^+)}{f(1)}\Big)=f'(1^+)-f(1)\log f(1)= -H(\rho)-\tau(\rho)\log \tau(\rho) \end{align*}
This justifies the second limit. For the uniform convergence on $D(\M)$, we note that
\[ \lim_{p\to 1}\frac{\tau(\rho^p)^{\frac{1}{p}}-\tau(\rho)}{p-1}=\lim_{p\to 1}\frac{\tau(\rho^p)-\tau(\rho)}{p-1}+\lim_{p\to 1}\frac{\tau(\rho^p)^{\frac{1}{p}}-\tau(\rho^p)}{p-1}\]
By mean value theorem,
\[\frac{x^{\frac{1}{p}} -x}{p-1}= -\frac{1}{p_0^2}x^{\frac{1}{p_0}}\ln x\]
for some $p_0\in (1,p)$. Note that on $D(\M)$, $\tau(\rho^p)\to 1$ uniformly. Then when $p\to 1$,
\[ \frac{\tau(\rho^p)^{\frac{1}{p}}-\tau(\rho^p)}{p-1}= -\frac{1}{p_0^2}\tau(\rho^p)^{\frac{1}{p_0}}\ln\tau(\rho^p) \to 0 \]
uniformly, which justifies the uniform convergence of the second limit. Then ii) follows from the uniform convergence of i).
\end{proof}

\subsection{Amalgamated $L_p$ norm and conditional entropy}
An important tool in our analysis is the amalgamated $L_p$-space introduced by Junge and Parcet \cite{junge2010mixed}.
Let  $\N\subset \M$ be a subalgebra, and let $\tau_\N$ be the trace of $\N$, which can be different with the tace $\tau_\M$ of $\M$. For $1\le p,q \le \infty $, fix $\frac{1}{r}=|\frac{1}{p}-\frac{1}{q}|$. Given $x\in \M$, the amalgamated $L_p^q$ norm is as follows: for $p\le q$,
\[ \norm{x}{L_p^q(\N\subset \M)}=\inf_{x=ayb } \norm{a}{L_{2r}(\N,\tau_\N)}\norm{y}{L_q(\M,\tau_\M)} \norm{b}{L_{2r}(\N,\tau_\N)}\pl;\]
where the infimum is for all factorization $x=ayb$ such that $a,b\in \N$ and $y\in \M$; for $p\ge q$
\[ \norm{x}{L_p^q(\N\subset \M)}=\sup_{\norm{\pl a\pl }{L_{2r}(\N)}=\norm{\pl b\pl }{L_{2r}(\N)}=1 } \norm{axb}{L_p(\M,\tau_\M)} \pl,\]
where the supremum is for all $a,b\in \N$ with $\norm{\pl a\pl }{L_{2r}(\N,\tau_\N)}=\norm{\pl b\pl }{L_{2r}(\N,\tau_\N)}=1$.
When $p=q$, the two definition are equivalent and $L_p^p(\N\subset \M)\cong L_p(\M,\tau_\M)$ isometrically.

For $x\ge 0$ , it suffices to consider $a=b>0$ in the above infimum (supremum). Then for $p\le q$,
\[ \norm{x}{L_p^q(\N\subset \M)}=\inf_{\sigma\in D_+(\N) } \norm{\sigma^{-\frac{1}{2r}}y \sigma^{-\frac{1}{2r}}}{L_q(\M,\tau_\M)} \pl;\]
for $q\le p$,
\[ \norm{x}{L_p^q(\N\subset \M)}=\sup_{\sigma\in D_+(\N) } \norm{\sigma^{\frac{1}{2r}}x \sigma^{\frac{1}{2r}}}{L_p(\M,\tau_\M)} \pl,\]
In particular, for $p=1,q=\infty$ and $p=\infty,q=1$ respectively, if $x\ge 0$,
\begin{align} &\norm{x}{L_1^\infty(\N\subset \M)}=\inf \{\lambda \pl | \pl x\le \lambda \sigma \text{ for some } \sigma\in D(\N)\}\\
&\norm{x}{L_\infty^1(\N\subset \M)}=\norm{E_N(x)}{\infty}
\end{align}
Here, $E_N:L_1(\M)\to L_1(\N)$ is the adjoint map of inclusion $\iota:\N\to\M$, defined as
\[  \tau_\M( x \rho )=  \tau_\N( xE_N(\rho) ) \pl, \pl \text{ for } x\in \N\pl, \rho\in L_1(\M)\]
Because of finite dimensions, we simply write $E_N:\M\to \N$.
Junge  and Parcet proved the following duality of amalgamated $L_p$ space
\[ L_{p}^{q}(\N\subset \M)^*=L_{p'}^{q'}(\N\subset \M)\pl, \]
where $\frac{1}{p}+\frac{1}{p'}=1$ and $\frac{1}{q}+\frac{1}{q'}=1$. Also, amalgamated $L_p$-spaces satisfies complex interpolation relation: for $0\le \theta\le 1$,
\[ L_{p_\theta}^{q_\theta}(\N\subset \M)=[L_{p_0}^{q_0}(\N\subset \M),L_{p_1}^{q_1}(\N\subset \M)]_\theta,\]
where $1\le p_j\le q_j\le \infty, \frac{1}{p_\theta}=\frac{1-\theta}{p_0}+\frac{\theta}{p_1}$ and  $ \frac{1}{q_\theta}=\frac{1-\theta}{q_0}+\frac{\theta}{q_1}$.
The next lemma shows the connection between amalgamated $L_p$-norms and entropy.

\begin{lemma}\label{lemma:diff}
If $\rho:[1,1+\eps)\in D(\M)$ satisfies $\lim_{p\to 1^+}\rho(p)=\rho$,
\begin{align} \lim_{p\to 1^+}\frac{1}{p-1}(\|\rho(p)\|_{L_{1}^{p}( \N\subset\M)}-1)\lel H(E_N(\rho))-H(\rho)\pl.\label{eq:diff}\end{align}
 \end{lemma}
\begin{proof} This is a modification of \cite[Theorem 17]{devetak2006multiplicativity}.  Without loss of generosity, we assume that $\tau_\N(e)\ge 1$ for any projections in $\N$. In fact, if for $\lambda,\mu>0$,
\[ \tilde{\tau}_\M=\lambda \tau_\M\pl, \tilde{\tau}_\N=\mu \tau_\N\pl, \pl \tilde{\rho} =\lambda^{-1}\rho  \pl,\]
both the entropy and $L_p$-norm only differ by a global constant,
\begin{align*} &H_{\tilde{\tau}}(\tilde{\rho})=H_{\tau}(\rho)-\log \lambda\pl,  H_{\tilde{\tau}_\N}(\tilde{E}_R(\tilde{\rho}))=H_{\tau_\N}(E_N(\rho))-\log \mu \\ &\norm{\tilde{\rho}}{L_p^q(\N\subset\M,\tilde{\tau})}=\mu^{1-\frac{1}{p}}\lambda^{\frac{1}{p}-1}\norm{\rho}{L_1^p(\N\subset\M,\tau)}\pl,\end{align*}
which match with \eqref{eq:diff}. Note that $\norm{\rho(p)}{L_{1}^{1}( \N\subset\M)}=\tau_\M(\rho(p))=1$, and
\[\norm{\rho(p)}{L_{1}^{p}( \N\subset\M)}=\inf_{\si\in D_+(\N)} \|\si^{-\frac{1}{2p'}}\rho(p)\si^{-\frac{1}{2p'}}\|_{p}=\inf_{\si\in D_+(\N)} \|\rho^{\frac12}(p) \si^{-\frac{1}{p'}}\rho^{\frac12}(p)\|_{p}\pl.\]
Denote \[\rho(p,\sigma)=\rho(p)^{\frac12} \si^{-\frac{1}{p'}}\rho(p)^{\frac12}\pl , \pl \hat{\rho}(p,\sigma)=\frac{\rho(p,\sigma)}{\tau_\M(\rho(p,\sigma))}\in D(\M) \pl.\]
It was proved in \cite{gao2020relative} that there exists an unique $\sigma$ attain the infimum in $\norm{\rho(p)}{L_{1}^{p}( \N\subset\M)}$, which we denote as $\sigma_p$. Namely,
\[ \norm{\rho(p)}{L_{1}^{p}( \N\subset\M)}=\norm{ \rho(p,\sigma_p)}{p}\pl.\]
By assumption $\displaystyle \inf_{e \text{ projection}}\tau_\N(e)\ge 1$, we have $\sigma^{-1}\ge 1\pl, \pl \forall \sigma \in D_+(\N)$. Then
\[ \rho(p)\le \rho(p,\sigma_p)\pl ,\pl\forall \pl p>1\pl.\]
On the other hand,
\begin{align*} 1=&\tau_\M(\rho(p))\le \tau_\M(\rho(p,\sigma_p))
\le  \tau_\M(1)^{1-\frac{1}{p}} \norm{\rho(p,\sigma_p)}{p}
\\ \le & \tau_\M(1)^{1-\frac{1}{p}} \norm{\rho(p,\frac{1}{\tau_\N(1)})}{p}
\le  \tau_\M(1)^{1-\frac{1}{p}} \tau_\N(1)^{1-\frac{1}{p}} \norm{\rho(p)}{p}\to 1
\end{align*}
Thus, $ \displaystyle \lim_{p\to 1^+}\rho(p,\sigma_p)=\lim_{p\to 1^+}\hat{\rho}(p,\sigma_p)=\rho$ in $L_1$-norm. Therefore,
 \begin{align}
 &\lim_{p\to 1^+}\frac{1}{p-1}(\|\rho(p)\|_{L_{1}^{p}( \N\subset\M)}-1)
 \\=&\lim_{p\to 1^+}\frac{ \|\rho(p,\sigma_p)\|_{p}-1}{p-1}\nonumber
 \\=&\lim_{p\to 1^+}\frac{ \|\rho(p,\sigma_p)\|_{p}-\|\rho(p,\sigma_p)\|_{1}}{p-1} +\frac{ \tau_\M(\rho(p,\sigma_p))-1}{p-1}\nonumber
 \\ \ge & \lim_{p\to 1^+}\tau_\M(\rho(p,\sigma_p))\frac{\|\hat{\rho}(p,\sigma_p)\|_{p}-1}{p-1}
 + \lim_{p\to 1^+}\inf_{\si}\frac{\tau_\M(\si^{-\frac{1}{p'}}\rho(p))-1}{p-1}\pl.
\end{align}
Using Lemma \ref{lemma:1} and $\tau_\M(\rho(p,\sigma))\to 1$, the first limit here converges to $-H(\rho)$. By H\"older inequality for $p<1$, the infimum in the second limit can be calculated
\[ \inf_{\si\in D_+(\N)}   \tau_\M(\si^{-\frac{1}{p'}}\rho(p))=\inf_{\si\in D_+(\N)}   \tau_\N(\si^{-\frac{1}{p'}}E_N(\rho(p)))=  \norm{E_N(\rho(p))}{\frac{p}{2p-1}}\pl. \]
Then by Lemma  \ref{lemma:1}, and chain rule, the second part converges to $H(E_N(\rho))$ as $E_N(\rho(p))\to E_N(\rho)$ . Hence we have
\begin{align}
 &\lim_{p\to 1^+}\frac{1}{p-1}(\|\rho(p)\|_{L_{1}^{p}( \N\subset\M)}-1)\ge H(E_N(\rho))-H(\rho)\label{eq:ge}\pl.
\end{align}
For the other direction,
\begin{align}
 &\lim_{p\to 1^+}\frac{1}{p-1}(\|\rho\|_{L_{1}^{p}( \N\subset\M)}-1)
\nonumber  \\ \ \le & \lim_{p\to 1^+}\frac{ \|\rho(p,E_N(\rho(p)))\|_{p}-1}{p-1}\nonumber
\nonumber  \\=& \lim_{p\to 1^+}\tau_\M(\rho(p,E_N(\rho(p))))\frac{\|\hat{\rho}(p,E_N(\rho(p)))\|_{p}-1}{p-1}
 + \lim_{p\to 1^+}\frac{\tau_\M( E_N(\rho(p))^{-\frac{1}{p'}}\rho(p))-1}{p-1}\nonumber
 \label{eq:le}
\end{align}
For the first limit, we note that by \cite{pimsner1986entropy}, there exists $C>0$ such that $\rho\le C E_N(\rho)$ for any $\rho\in \M_+$. Then
\begin{align*}&\rho(p)\le \rho(p,E_N(\rho(p)))=\rho(p)^{\frac12} E_N(\rho(p))^{-\frac{1}{p'}}\rho(p)^{\frac12}\le C^{\frac{1}{p'}} \rho(p)^{\frac{1}{p}}
\\ 1=&\tau(\rho(p))\le \tau(\rho(p,E_N(\rho(p))))\le C^{\frac{1}{p'}} \tau(\rho(p)^{\frac{1}{p}})\to 1
\end{align*}
Thus $\rho(p,E_N(\rho(p)))\to \rho$ in $L_1$ norm, which implies
\[\lim_{p\to 1^+}\tau_\M(\rho(p,E_N(\rho(p))))\frac{\|\hat{\rho}(p,E_N(\rho(p)))\|_{p}-1}{p-1}=-H(\rho)\pl.\]
For the second limit, we note that $\displaystyle\lim_{p\to 1^+} \norm{E_N(\rho(p))-E_N(\rho)}{1}\le \lim_{p\to 1^+} \norm{\rho(p)-\rho}{1}=0$. Then by Lemma \ref{lemma:1} again and chain rule,
\begin{align*} &\lim_{p\to 1^+}\frac{\tau_\M( E_N(\rho(p))^{-\frac{1}{p'}}\rho(p))-1}{p-1}= \lim_{p\to 1^+}\frac{\tau_\M( E_N(\rho(p))^{\frac{1}{p}})-1}{p-1}=-H(E_N(\rho)) \qedhere\end{align*}
\end{proof}

\begin{exam}{\rm \label{exam:s1p} Consider the matrix algebra $(\M,\tau_\M)=(\bB(H_A\ten H_B),\tr_{AB}),(\N,\tau_\N)=(\bB(H_B),\tr_B )$, $\N\cong\mathbb{C}1\ten\bB(H_B)\subset \bB(H_A\ten H_B)$, the $L_1^p$-norm for positive $X_{AB}\in \bB(H_A\ten H_B)_+$  is
\begin{align}
\norm{X_{AB} }{S_1(H_B,S_p(H_A))}=\inf_{\sigma\in \bB(H_B)} \tr( |(1\ten \sigma^{-\frac{1}{2p'}})  \rho_{AB} (1\ten \sigma^{-\frac{1}{2p'}})|^p)^{\frac1p}\pl ,
\end{align}
where the infimum is for all density operator $\sigma\in \bB(H_B)$.
This case was introduced by  Pisier \cite{pisier1998non}.
It was proved in \cite[Theorem 17]{devetak2006multiplicativity} that for density operator $\rho_{AB}$,
\begin{align}\lim_{p\to 1^+}\frac{1}{p-1}{(\|\rho\|_{S_1(H_B,S_p(H_A))}-1)}\lel H(\rho_{B})-H(\rho_{AB}):=-H(A|B)_\rho\pl, \label{eq:cdentropy}\end{align}
where $H(A|B)$ is called conditional entropy. Moreover,
\[ H_p(A|B)_\rho:=\frac{p}{p-1}\log \norm{\rho_{AB} }{S_1(H_B,S_p(H_A))}\]
is the sandwiched R\'enyi $p$-conditional entropy \cite{muller2013quantum}. In particular, \eqref{eq:cdentropy} implies
\begin{align}\lim_{p\to 1^+}H(A|B)_\rho=-H(A|B)_\rho\pl. \label{eq:cdentropy1}\end{align}
}\end{exam}

\subsection{Kosaki $L_p$-norm and relative entropy}
Given an invertible positive operator $\sigma\in \M_+$, Kosaki \cite{kosaki1984applications} introduced the following weighted $L_p$-space:
\[ \norm{x}{\sigma,p}= \tau(|\sigma^{\frac{1}{2p}}x\sigma^{\frac{1}{2p}}|^p)^{\frac{1}{p}}\pl.\]
We denote $L_p(\M,\sigma)$ as the space for the above norm. It is known that Kosaki $L_p$-space also satisfies complex interpolation space: for $0\le \theta\le 1$,
\[ L_{p_\theta}(\M,\sigma)=[L_{p_0}(\M,\sigma),L_{p_1}(\M,\sigma)]_\theta\pl, \]
where $1\le p_0\le q_1\le \infty$, and $\frac{1}{p_\theta}=\frac{1-\theta}{p_0}+\frac{\theta}{p_1}$.

Given a density operator $\rho\in \cD(\M)$, the relative entropy with respect to $\sigma$ is defined as
\[ D(\rho||\sigma)=\tau(\rho\log \rho-\rho\log\sigma)\pl.\]
Note that the above definition is independent of trace $\tau$, only depends on the state $\rho$ and $\sigma$. The relation to  Kosaki $L_p$-norm is as follows.

\begin{lemma}\label{lemma:dp}Given $\sigma\in \M_+$, for $\rho\in D(\M)$, we have uniform convergence
\begin{align*}
\lim_{p\to 1}\frac{\norm{\si^{-\frac{1}{2}}\rho\si^{-\frac{1}{2}}}{\si, p}^p-1}{p-1}=\lim_{p\to 1}\frac{\norm{\si^{-\frac{1}{2}}\rho\si^{-\frac{1}{2}}}{\si, p}-1}{p-1}= D(\rho||\si)
\end{align*}
If $\rho:[1,1+\eps)\to D(\M)$ satisfies $\displaystyle\lim_{p\to 1^+}\rho(p)= \rho$, then
\begin{align*}
\lim_{p\to 1}\frac{\norm{\si^{-\frac{1}{2}}\rho(p)\si^{-\frac{1}{2}}}{\si, p}-1}{p-1}=D(\rho||\si)
\end{align*}
\end{lemma}
\begin{proof}Fix $p'=\frac{p}{p-1}$.
We denote
\[ \rho_p=\si^{-\frac{1}{2p'}}\rho\si^{-\frac{1}{2p'}}, \hat{\rho}_p=\frac{\rho_p}{\tau_\M(\rho_p)}\in D(\M)\pl.\]
Because $\sigma$ is invertible, $\rho(p)$ is continuous with respect to $p$ and $\rho(1)=\hat{\rho}(1)=\rho$. By Lemma \ref{lemma:1},
\begin{align}
 &\lim_{p\to 1^+}\frac{1}{p-1}(\norm{\si^{-\frac{1}{2}}\rho\si^{-\frac{1}{2}}}{\si, p}^p-1)
 \\=&\lim_{p\to 1^+}\frac{\norm{\rho_p}{p}^p-1}{p-1}\nonumber
 \\=&\lim_{p\to 1^+}\frac{\norm{\rho_p}{p}^p-\norm{\rho_p}{1}}{p-1}+\lim_{p\to 1^+}\frac{\norm{\rho_p}{1}-1}{p-1}\nonumber
 \\=&\lim_{p\to 1^+}\tau(\si^{-\frac{1}{p'}}\rho)\frac{\norm{\hat{\rho}_p}{p}-1}{p-1}+\lim_{p\to 1^+}\frac{\tau(\si^{-\frac{1}{p'}}\rho)-1}{p-1}\nonumber
\\=&\tau(\rho\log \rho)-\tau(\rho \log \sigma)\nonumber
\\=& D(\rho||\sigma)\nonumber\pl,
\end{align}
where both limit in the above calculation are uniform. The second assertion follows from the uniform continuity of $\rho\mapsto D(\rho||\sigma)$ ($\sigma$ is invertible and fixed).
\end{proof}

\begin{rem}{\rm
$D_p(\rho||\sigma)=\frac{p}{p-1}\log \norm{\sigma^{-\frac{1}{2}}\rho \sigma^{-\frac{1}{2}}}{\sigma,p}$ is called Sandwiched R\'enyi relative entropy \cite{muller2013quantum, wilde2014strong}. The above argument shows
\[\lim_{p\to 1^+}D_p(\rho||\sigma)=D(\rho||\sigma)\]
}
\end{rem}

We will also need weighted amalgamated $L_p$-space. Let $\N\subset \M$ be a subalgebra. Recall that a map $E_N^\dagger:\M\to \N$ is called a conditional expectation if  $E_N^\dagger$ is complete positive map satisfying $E_N^\dagger\circ E_N^\dagger=E_N^\dagger$. Given a conditional expectation $E_N^\dagger$, $\N$ admits a canonical trace $\tau_\N=\tau_\M|_\N$, whose density operator w.r.t $\tau_\M$ is $\sigma_{\tr}=E_N(1)$, where $E_N$ is the adjoint of  $E_N^\dagger $. ($\sigma_{\tr}\in \N'$, see \cite{bardet2017estimating,gao2022complete}). We have
\[ E_N( \sigma^{\frac{1}{2}}_{\tr }x\sigma^{\frac{1}{2}}_{\tr })=\sigma^{\frac{1}{2}}_{\tr }E_N^*( x)\sigma^{\frac{1}{2}}_{\tr }\pl, \]
and the chain rule for relative entropy \cite{junge2023stability},
\begin{align}
D(\rho|| E_N(\rho))=D(\rho||\sigma)-D(E_N(\rho)||\sigma)\label{eq:chainrule}\pl,
\end{align}
which holds for any $\sigma$ satisfying $E_N(\sigma)=\sigma$.

For $1\le p\le \infty \pl, \frac{1}{p}+\frac{1}{p'}=1$, we define the norm
\[ \norm{x}{L_1^p(\N\subset \M,\sigma_{\tr})}:= \inf_{x=ayb} \norm{a}{L_{2p'}(\N,\sigma_{\tr})} \norm{ y }{L_{p}(\M,\sigma_{\tr})}\norm{b}{L_{2p'}(\N,\sigma_{\tr} )}\pl.\]
where the infimum is over all factorization $x=ayb$ satisfying $a,b\in \N$.
This space also satisfies complex interpolation: for $\theta\in [0,1]$,
\[ L_{1}^{p_\theta}(\N\subset \M, \sigma_{\tr})=[L_{1}^{p_0}(\N\subset \M, \sigma_{\tr}),L_{1}^{p_1}(\N\subset \M, \sigma_{\tr})]\]
where $\frac{1}{p_\theta}=\frac{1-\theta}{p_0}+\frac{\theta}{p_1}, 1\le p_0\le p_1\le \infty$.

\begin{lemma}\label{lemma:dp1}
If $\rho:[1,1+\eps)\to D(\M) , p \mapsto\rho(p)$ satisfies  $ \displaystyle \lim_{p\to 1^+} \rho(p)=\rho$, then
\begin{align*}
\lim_{p\to 1}\frac{\norm{\sigma_{\tr}^{-\frac{1}{2}}\rho(p) \sigma_{\tr}^{-\frac{1}{2}}}{L_1^p(\N\subset \M,\sigma_{\tr})}-1}{p-1}=D(\rho||\si_{\tr})-D(E_N(\rho)||\sigma_{\tr})\pl.
\end{align*}
\end{lemma}
\begin{proof} Let  $\gamma\in \N_+$ such that $\tau_\M(\gamma\sigma_{\tr})=\tau_\M(\gamma)=1$ .
Denote
\[\rho(p,\gamma)=\rho(p)^{\frac12}\sigma_{\tr}^{-\frac{1}{2p'}}\gamma^{-\frac{1}{p'}}\sigma_{\tr}^{-\frac{1}{2p'}} \rho(p)^{\frac12} \pl, \pl \hat{\rho}(p,\gamma)=\frac{\rho(p,\gamma) }{\tau_\M(\rho(p,\gamma))}\]
By definition
\begin{align*} \norm{\sigma_{\tr}^{-\frac{1}{2}}\rho(p) \sigma_{\tr}^{-\frac{1}{2}}}{L_{1}^{p}( \N\subset\M)}=
&\inf_{\gamma} \norm{ \gamma^{-\frac{1}{2p'}}\sigma_{\tr}^{-\frac{1}{2}}\rho(p)\sigma_{\tr}^{-\frac{1}{2}}  \gamma^{-\frac{1}{2p'}}}{L_{p}(\M,\sigma_{\tr})}
\\
=&\inf_{\gamma} \norm{ \gamma^{-\frac{1}{2p'}}\sigma_{\tr}^{-\frac{1}{2p'}}\rho(p)\sigma_{\tr}^{-\frac{1}{2p'}}  \gamma^{-\frac{1}{2p'}}}{p}\\
=&\inf_{\gamma} \norm{\rho(p,\gamma)}{p}=\norm{\rho(p,\gamma_p)}{p}
\end{align*}
Since in finite dimensions, we can assume the infimum is attained by some $\gamma_p\in D(\N)$.
Similar to the proof of Lemma \ref{lemma:diff}, we can assume $\displaystyle \inf_{e \text{ projection}}\tau_\M(e)\ge 1$.
Then for all $\gamma\in \N_+$ satisfying $\tau_\M(\gamma\sigma_{\tr})=\tau_\M(\gamma)=1$, we have  $\sigma_{\tr}^{-1}\gamma^{-1}\ge 1,\gamma^{-1}\ge 1$ ($\sigma_{\tr}$ and $\sigma$ commute). Then
\[ \rho(p) \le \rho(p)^{\frac{1}{2}} \sigma_{\tr}^{-\frac{1}{p'}}\rho(p)^{\frac{1}{2}}\le \rho(p,\gamma_p)\pl ,\pl\forall \pl p>1\pl.\]
On the other hand,
\begin{align*} &1=\tau(\rho(p))\le \tau(\rho(p,\gamma_p))
\le  \tau_\M(1)^{1-\frac{1}{p}} \norm{\rho(p,\gamma_p)}{p}
\\ \le & \tau_\M(1)^{1-\frac{1}{p}} \norm{\rho(p,\frac{1}{\tau_\M(1)})}{p}
\le  \tau_\M(1)^{2-\frac{2}{p}} \norm{\sigma_{\tr}^{-\frac{1}{2p'}}\rho(p) \sigma_{\tr}^{-\frac{1}{2p'}}}{p}\to 1
\end{align*}
Then\[
\lim_{p\to 1^+}\rho(p,\gamma_p)=\lim_{p\to 1^+}\hat{\rho}(p,\gamma_p)=\lim_{p\to 1^+}\rho^{\frac{1}{2}} \sigma_{\tr}^{-\frac{1}{p'}}\rho^{\frac{1}{2}}= \rho\] in $L_1$-norm. This implies
\begin{align*}\norm{\rho^{\frac{1}{2}}(p)\gamma_p^{-\frac{1}{2p'}}-\rho^{1/2}(p)}{2}\le& \norm{\sigma_{\tr}^{\frac{1}{2p'}}}{\infty} \norm{\rho^{\frac{1}{2}}(p)\gamma_p^{-\frac{1}{2p'}}\sigma_{\tr}^{-\frac{1}{2p'}}-\rho^{1/2}(p)\sigma_\tr^{-\frac{1}{2p'}}}{2}\\
\le& \norm{\sigma_{\tr}^{\frac{1}{2p'}}}{\infty} \tau(\rho(p)\gamma_p^{-\frac{1}{p'}}\sigma_\tr^{-\frac{1}{p'}}-2\rho(p)\gamma_p^{-\frac{1}{2p'}}\sigma_\tr^{-\frac{1}{p'}}+\rho(p)\sigma_\tr^{-\frac{1}{p'}})
\to 0
\end{align*}
Hence, $\displaystyle \lim_{p\to 1^+}\gamma_p^{-\frac{1}{2p'}}\rho(p)\gamma_p^{-\frac{1}{2p'}}=\rho(p) $. Denote\[ \rho_p(\gamma)=\gamma^ {-\frac{1}{2p'}}\rho(\rho)\gamma^{-\frac{1}{2p'}}\pl, \pl \tilde{\rho}_p(\gamma)=\frac{\rho_p(\gamma)}{\tau_\M(\rho_p(\gamma))}\pl. \]We have
 \begin{align}
 &\lim_{p\to 1^+}\frac{1}{p-1}(\|\sigma_\tr^{-\frac{1}{2}}\rho(p) \sigma_\tr^{-\frac{1}{2}}\|_{L_{1}^{p}( \N\subset\M,\sigma_\tr)}-1)\nonumber
 \\=&\lim_{p\to 1^+}\frac{ \|\sigma_\tr^{-\frac{1}{2}}\rho_p(\gamma_p)\sigma_\tr^{-\frac{1}{2}}\|_{p,\sigma_\tr}-1}{p-1}\nonumber
 \\=&\lim_{p\to 1^+}\frac{ \|\sigma_\tr^{-\frac{1}{2}}\rho_p(\gamma_p)\sigma_\tr^{-\frac{1}{2}}\|_{p}-\|\sigma_\tr^{-\frac{1}{2}}\rho_p(\gamma_p)\sigma_\tr^{-\frac{1}{2}}\|_{1,\sigma_\tr}}{p-1} +\frac{ \tau_\M(\rho_p(\gamma_p))-1}{p-1}\nonumber
 \\ \ge & \lim_{p\to 1^+}\tau_\M(\rho_p(\gamma_p))\frac{\|\sigma_\tr^{-\frac{1}{2}}\rho_p(\gamma_p)\sigma_\tr^{-\frac{1}{2}}\|_{p}-1}{p-1}
 + \lim_{p\to 1^+}\inf_{\gamma}\frac{\tau_\M(\gamma^{-\frac{1}{p'}}\rho(p))-1}{p-1}\pl.
\end{align}
Here, the first limit converges to  $D(\rho||\sigma_{\tr})$ by Lemma \ref{lemma:dp}. The infimum in the second limit can be calculated
\begin{align*} \inf_{\gamma}\tau_\M(\gamma^{-\frac{1}{p'}}\rho(p))=& \inf_{\gamma}\tau_\M(\gamma^{-\frac{1}{p'}}E_N(\rho(p)))\\
=& \inf_{\gamma}\tau_\M(\gamma^{-\frac{1}{p'}} \sigma_{tr}^{\frac{1}{2}}E_N^*(\sigma_{tr}^{-\frac{1}{2}}\rho(p) \sigma_{tr}^{-\frac{1}{2}})) \sigma_{tr}^{\frac{1}{2}})\\= &\inf_{\gamma}\sigma_{tr}( \gamma^{-\frac{1}{p'}} E_N^*(\sigma_{tr}^{-\frac{1}{2}}\rho(p) \sigma_{tr}^{-\frac{1}{2}}))=\norm{ \sigma_{tr}^{-\frac{1}{2}}E_N(\rho (p) )\sigma_{tr}^{-\frac{1}{2}}}{\frac{p}{2p-1},\sigma_{tr}}
 \end{align*}
 Note that $\sigma_\tr$ is a trace on $\N$ , and $\displaystyle \lim_{p\to 1^+}E_N(\rho (p) )= E_N(\rho)$. Then by Lemma \ref{lemma:dp}
\begin{align*} \lim_{p\to 1^+}\frac{\norm{\sigma_{\tr}^{-\frac{1}{2}}E_N(\rho (p)) \sigma_{\tr}^{-\frac{1}{2}} }{\frac{p}{2p-1},\sigma_{\tr}}-1}{p-1}=&-\sigma_{\tr}( E_N(\rho) \log E_N(\rho ))=-D(E_N(\rho)||\sigma_{\tr}) \pl,\end{align*}
where we use the fact $D(\rho||\sigma)$ is independent of trace.
For the other direction, we denote $\rho_N(p)=E_N(\rho(p))$ and take $\hat{\gamma} _p=\sigma_\tr^{-\frac{1}{2}}E_N(\rho(p))\sigma_\tr^{-\frac{1}{2}}, \rho_p=\frac{\hat{\gamma}_p^{-\frac{1}{2p'}}\rho(p)  \hat{\gamma}_p^{-\frac{1}{2p'}}}{\tau_\M(\hat{\gamma}_p^{-\frac{1}{2p'}}\rho(p)  \hat{\gamma}_p^{-\frac{1}{2p'}})}$,
\begin{align}
 &\lim_{p\to 1^+}\frac{1}{p-1}(\|\sigma_\tr^{-\frac{1}{2}}\rho(p) \sigma_\tr^{-\frac{1}{2}}\|_{L_{1}^{p}( \N\subset\M,\sigma_\tr)}-1)
\nonumber  \\ \ \le & \lim_{p\to 1^+}\frac{ \| \sigma_\tr^{-\frac{1}{2}}\hat{\gamma}_p^{-\frac{1}{2p'}}\rho(p)  \hat{\gamma}_p^{-\frac{1}{2p'}}\sigma_\tr^{-\frac{1}{2}}\|_{p,\sigma_\tr}-1}{p-1}\nonumber
\nonumber  \\=& \lim_{p\to 1^+}\tau_\M(\hat{\gamma}_p^{-\frac{1}{2p'}}\rho(p)  \hat{\gamma}_p^{-\frac{1}{2p'}})\frac{\|\sigma_\tr^{-\frac{1}{2}}\rho_p\sigma_\tr^{-\frac{1}{2}}\|_{p,\sigma_\tr}-1}{p-1}
 + \lim_{p\to 1^+}\frac{\tau_\M(\hat{\gamma}_p^{-\frac{1}{p'}}\rho(p) )-1}{p-1}\nonumber
 \\=& D(\rho||\sigma_\tr)-D(E_N(\rho)||\sigma_\tr)\label{eq:le}
\end{align}
Here, for the first limit follows from Lemma  \ref{lemma:dp} and
\begin{align*}&\norm{\rho^{\frac{1}{2}}(p)\hat{\gamma}_p^{-\frac{1}{2p'}}-\rho^{\frac{1}{2}}(p)}{2}\\ \le& \norm{\sigma_{\tr}^{\frac{1}{2p'}}}{\infty} \norm{\rho^{\frac{1}{2}}(p)\hat{\gamma}_p^{-\frac{1}{2p'}}\sigma_{\tr}^{-\frac{1}{2p'}}-\rho^{\frac{1}{2}}(p)\sigma_\tr^{-\frac{1}{2p'}}}{2}\\
\le& \norm{\sigma_{\tr}^{\frac{1}{2p'}}}{\infty} \tau(\rho(p)\hat{\gamma}_p^{-\frac{1}{p'}}\sigma_\tr^{-\frac{1}{p'}}-2\rho(p)\hat{\gamma}_p^{-\frac{1}{2p'}}\sigma_\tr^{-\frac{1}{p'}}+\rho(p)\sigma_\tr^{-\frac{1}{p'}})
\\= &\norm{\sigma_{\tr}^{\frac{1}{2p'}}}{\infty} \tau(\rho(p)E_N(\rho(p))^{-\frac{1}{p'}}-2\rho(p)E_N(\rho(p))^{-\frac{1}{p'}}\sigma_\tr^{-\frac{1}{2p'}}+\rho(p)\sigma_\tr^{-\frac{1}{p'}})
\to 0,
\end{align*}
where we use the fact $ \rho(p)\le C E_N(\rho(p))$ for some finite $C$ (see \cite{gao2022complete}).
 For the second limit, we have
\begin{align*}\tau(\hat{\gamma}_p^{-\frac{1}{p'}} \rho  )=&\tau(\hat{\gamma}_p^{-\frac{1}{p'}}\sigma_{\tr}^{\frac{1}{2}} (\sigma_{\tr}^{-\frac{1}{2}}\rho(p)  \sigma_{\tr}^{-\frac{1}{2}}) \sigma_{\tr}^{\frac{1}{2}})\\=&\tau(\hat{\gamma}_p^{-\frac{1}{p'}}\sigma_{\tr}^{\frac{1}{2}} E_N(\sigma_{\tr}^{-\frac{1}{2}}\rho(p)  \sigma_{\tr}^{-\frac{1}{2}}) \sigma_{\tr}^{\frac{1}{2}})
\\=&\tau(\hat{\gamma}_p^{-\frac{1}{p'}}\sigma_{\tr}^{\frac{1}{2}} E_N(\sigma_{\tr}^{-\frac{1}{2}}\rho(p)  \sigma_{\tr}^{-\frac{1}{2}}) \sigma^{\frac{1}{2}})
\\=&\tau(\hat{\gamma}_p^{-\frac{1}{p'}}\sigma_{\tr}^{\frac{1}{2}} \hat{\gamma}_p \sigma_{\tr}^{\frac{1}{2}})
\\=& \sigma_{\tr}(\hat{\gamma}_p^{\frac{1}{p}} )= \norm{\hat{\gamma}_p}{\frac{1}{p},\sigma_\tr}^{1/p}\pl,\\
\lim_{p\to 1^+}\frac{\tau_\M(\hat{\gamma}_p^{-\frac{1}{p'}}\rho(p))-1}{p-1}=& \lim_{p\to 1^+}\frac{\norm{\sigma_{\tr}^{\frac{1}{2}} E_N(\rho(p))\sigma_{\tr}^{-\frac{1}{2}}}{\frac{1}{p},\sigma_\tr }^{\frac{1}{p}}-1}{p-1}\\ =& -\sigma_{\tr}( \hat{\gamma}_p \log \hat{\gamma}_p)= -D(E_N(\rho)||\sigma_{\tr})
\end{align*}
where we used again Lemma \ref{lemma:dp} and $ E_N(\rho(p))\to E_N(\rho)$ as $p\to 1$.
\end{proof}

\section{Generalized Strong Sub-additivity of quantum channels}
Let $(\M,\tau_\M)$ and $(\N,\tau_\N)$ be two finite dimensional von Neumann algebras. We say a linear map $\Phi:\M\to \N$ is positive if $\Phi(\M_+)\subset \N_+$ ; completely positive, if for any matrix algebra $\bM_n$, $\Phi\ten\id_{\bM_n}$ is positive; $\Phi$ is trace preserving, if for any  $\rho\in\M$ , $\tau_\N(\Phi(\rho))=\tau_\M(\rho)$. A completely positive trace preserving (CPTP) map is called a quantum channel, which send density operators to density operators. The adjoint map $\Phi^\dagger:\N\to\M$ is completely positive and unital $\Phi^\dagger(1)=1$ (UCP). A special case is when $\N\subset\M$ is a subalgebra, the embedding map $\iota_\N: \N\to\M$ is clearly a UCP map. It adjoint map $E_\N: \M\to\N$ is a quantum channel. For example,  $\bB(H_B)\cong \mathbb{C}1\ten \bB(H_B)\subset \bB(H_B\ten H_A)$, the partial trace map $\tr_A=\tr\ten \id_B:\bB(H_B\ten H_A)\to  \bB(H_B)$ is CPTP.
\begin{theorem}\label{thm:hssa}
Let $\cA,\cB,\cM$ and $\cR$ be finite dimensional von Neumann algebras with traces denoted as $\tau_\A,\tau_\B,\tau_\M$ and $\tau_\R$ respectively. Assume that $\cR\subset \cA$ is a subalgebra, and denote $E_R$ as the adjoint map of the embedding. Given two quantum channel map $\Phi_A:\M\to \A$ and $\Phi_B:\M\to \B$,  for any density operator $\rho\in\cM$, we have
\begin{align}H(\Phi_A(\rho))+H(\Phi_B(\rho))\ge H(\rho)+H(E_R\circ \Phi_A(\rho))+\log \frac{1}{c}\pl,\end{align}
where the constant $c$ is
\[c=\sup \{\tau_\M(\Phi_A^\dag(a)\Phi_B^\dag(b)) \pl |\pl a\in \A_+,  E_R(a)= 1\pl,b\in D(\B) \}\pl,\]
\end{theorem}
\begin{proof} Fix a density operator $b\in D(\B)$. For $0\le \Re(z)\le 1$, we define an analytic family of map $T_z:\M\to \A$
\[T_z(\rho)= \Phi_A\Big(\Phi_B^\dag(b)^{\frac{1-z}{2}}\rho\Phi_B^\dag(b)^{\frac{1-z}{2}}\Big)  \pl.\]
For $z=it$, by the duality $L_1^\infty(\R\subset\A)^*=L_\infty^1(\R\subset\A)$
\begin{align}
\norm{T_{it}:L_\infty(\M)\to L_1^\infty(\R\subset\A)}{}&= \sup_{\norm{\pl\rho\pl }{\infty}=1}\sup_{\norm{\pl a\pl}{ L_\infty^1(\R\subset\A)}=1} |\tau_\M\Big(a\Phi_A\big(\Phi_B^\dag(b)^{\frac{1-it}{2}}\rho\Phi_B^\dag(b)^{\frac{1-it}{2}}\big)
\Big)|\nonumber\\
&= \sup_{\norm{\pl \rho\pl}{\infty}=1}\sup_{\norm{\pl a\pl}{L_\infty^1}=1} |\tau_M\Big(\Phi_B^\dag(b)^{\frac{1}{2}}\rho\Phi_B^\dag(b)^{\frac{1}{2}}\Phi_A^\dag(
a)\Big)|\nonumber \\
&= \sup_{\norm{\pl a \pl}{L_\infty^1}=1} \norm{\Phi_B^\dag(b)^{\frac{1}{2}}\Phi_A^\dag(
a)\Phi_B^\dag(b)^{\frac{1}{2}}}{L_1(\M)} \nonumber\\
&= \sup_{a\ge 0,\pl  E_R(a)\le 1} \tau_\M\Big(\Phi_B^\dag(b)^{\frac{1}{2}}\Phi_A^\dag(
a)\Phi_B^\dag(b)^{\frac{1}{2}}\Big) \label{eq:1}\\
&= \sup_{a\ge 0,\pl  E_R(a)\le 1} \tau_\M\Big(\Phi_B^\dag(b)\Phi_A^\dag(
a)\Big):=c(b)\pl. \nonumber
\end{align}
Here, equality \eqref{eq:1} uses the fact that
\[ S:L_\infty^1(\R\subset\A)\to L_1(\M)\pl, \pl  a\mapsto \Phi_B^\dag(b)^{\frac{1}{2}}\Phi_A^\dag(
a)\Phi_B^\dag(b)^{\frac{1}{2}}\]
is completely positive, then the map norm can be attained by positive elements \cite[Theorem 13]{devetak2006multiplicativity}. By definition, $c=\displaystyle \sup_{b\in D(\B)}c(b)$.
For $z=1+it$,
\begin{align*}\norm{T_{1+it}:L_1(\M)\to L_1(\A)}{}&= \sup_{\norm{\pl \rho\pl }{1}=1} \norm{\Phi_A(\Phi_B^\dag(b)^{\frac{-it}{2}}\rho\Phi_B^\dag(b)^{\frac{-it}{2}})}{L_1(\A)}\\
&\le\sup_{\norm{\pl \rho\pl }{ 1}=1}
\norm{\Phi_A(\rho)}{L_1(\A)}\\
&= \norm{\Phi_A: L_1(\M)\to L_1(\A)}{}=1\pl,
\end{align*}
because $\Phi_A$ is positive and trace preserving.
By interpolation (see  \cite{bergh2012interpolation}), we know for any  $b\in D(\B)$, \[ \norm{T_p: L_p(\M)\to L_p(\A)}{}\le c(b)^{1-\frac{1}{p}}\pl.\]
Then for any $\rho\in D(\M)$
\begin{align} \norm{\Phi_A(\Phi_B^\dag(b)^{\frac{1}{2p'}}\rho\Phi_B^\dag(b)^{\frac{1}{2p'}})}{L_1^p(\R\subset\A)}\le \norm{\rho}{L_p(\M)} c(b)^{1-\frac{1}{p}}\pl. \label{eq:interpolation}\end{align}
Denote \begin{align*}
&\omega(p)=\Phi_A\Big(\Phi_B^\dag(b)^{\frac{1}{2p'}}\rho\Phi_B^\dag(b)^{\frac{1}{2p'}}\Big) \pl, \pl \hat{\omega}(p)=\frac{\omega(p)}{\tau_A(\omega(p))}
\end{align*}
Thus, we have $\omega(1)=\hat{\omega}(1)=\Phi_A(\rho)$ and
 \begin{align}\label{eq:p}\lim_{p\to 1^+} \frac{\norm{\omega(p)}{L_1^p(\R\subset\A)}-1}{p-1}\le \lim_{p\to 1^+} \frac{\norm{\rho}{L_p(\M)} c(b)^{1-\frac{1}{p}}-1}{p-1}\pl.\end{align}
Since $\Phi_A$ is trace preserving, $ \tau_\M(\rho)=\tau_\A(\Phi_A(\rho))=1$. We apply Lemma \ref{lemma:1} for the right hand side of \eqref{eq:p},
\begin{align} &\lim_{p\to 1^+} \frac{\norm{\rho}{L_p(\M)}c(b)^{1-\frac{1}{p}}-1}{p-1}\\ =&\lim_{p\to 1^+} \norm{\rho}{L_p(\M)}\frac{(c(b)^{1-\frac{1}{p}}-1)}{p-1}+\lim_{p\to 1^+} \frac{\norm{\rho}{L_p(\M)}-1}{p-1}=\ln c(b) -H(\rho)\pl.\end{align}
For the left hand side of \eqref{eq:p},
\begin{align*}
&\lim_{p\to 1^+} \frac{\norm{\omega(p)}{L_1^p(\R\subset\A)}-1}{p-1}
\\=&\lim_{p\to 1^+}\frac{\tau_A(\omega(p))\norm{\hat{\omega}(p)}{L_1^p(\R\subset\A)}-1}{p-1}
\\ = &\lim_{p\to 1^+}\tau_A(\omega(p))\frac{\norm{\hat{\omega}(p)}{L_1^p(\R\subset\A)}-1}{p-1} +\lim_{p\to 1^+}\frac{\tau_A(\omega(p))-1}{p-1}
\end{align*}
By Lemma \ref{lemma:diff}, the first term is
\[ \lim_{p\to 1^+}\tau_A(\omega(p))\frac{\norm{\hat{\omega}(p)}{L_1^p(\R\subset\A)}-1}{p-1}=-H(\Phi_A(\rho))+H(E_R\circ \Phi_A(\rho))\pl.\]
For the second term, because again $\Phi_A$ is trace preserving, we have
\begin{align} \tau_A\Big(\Phi_A\big(\Phi_B^\dag(b)^{\frac{1}{2p'}}\rho\Phi_B^\dag(b)^{\frac{1}{2p'}}\Big)\Big)=\tau_\M(\Phi_B^\dag(b)^{\frac{1}{p'}}\rho)\ge \tau_\M(\Phi_B^\dag(b^{\frac{1}{p'}})\rho)=\tau_\B(b^{\frac{1}{p'}}\Phi_B(\rho)) \pl.\end{align}
Here we use the operator convexity of $f(x)=x^{\frac{1}{p'}}$. Take $b=\Phi_B(\rho)$,
\begin{align} \lim_{p\to 1^+}\frac{\tau_A(\omega(p))-1}{p-1}\ge \lim_{p\to 1^+}\frac{\tau_\B(\Phi_B(\rho)^{2-\frac{1}{p}})-1}{p-1}=-H(\Phi_B(\rho))\pl.\end{align}
Combining all the steps above, we have
\begin{align*} &\log c-H(\rho)\ge -H(\Phi_B(\rho))+H(E_R\circ \Phi_A(\rho))
-H(\Phi_A(\rho))\pl. \qedhere\end{align*}
\end{proof}

\begin{rem}{\rm \label{rem:constant} In fact, we proved
\begin{align}H(\Phi_A(\rho))+H(\Phi_B(\rho))\ge H(\rho)+H(E_\R\circ \Phi_A(\rho))+\log \frac{1}{c(\rho)}\pl,\end{align}
where $c(\rho)$ is a local constant depending on $\rho$
\begin{align*}c(\rho)=&\sup\{ \tau_\M\Big(\Phi_B^\dag(\Phi_B(\rho))\Phi_A^\dag(
a)\Big)\pl |\pl a\in \A_+\pl, E_R(a)\le 1\}
\\ =&\norm{\Phi_A\circ\Phi_B^\dag\circ\Phi_B(\rho)}{L_1^\infty(\A\subset \R)}
\end{align*}
while the global constant $c$ in above theorem is
\[ c=\norm{\Phi_A\circ\Phi_B^\dag: L_1(\B)\to L_1^\infty(\A\subset \R)}{}\ge c(\rho)\pl.\]}
\end{rem}

\begin{exam}{\rm Consider a simple case : $\cR=\mathbb{C}1$ is trivial subalgebra, Theorem \ref{thm:hssa} becomes
\[ H(\Phi_A(\rho))+H(\Phi_B(\rho))\ge H(\rho)+\log\frac{1}{c} \]
where the constant
\[c=\sup_{a\in D(\A)\pl, \pl b\in D(\B)} \tau_\M(\Phi^\dagger_A(a)\Phi^\dagger_B(b))\pl.\]
This constant is a noncommutative analog of maximum overlap of two measurements in Frank-Lieb uncertainty relation \cite{frank2013extended}.
This case can also be derived from quantum Brascamp-Lieb duality by Berta, Sutter  and Walter  \cite{berta2019quantum}. Actually, they obtained a stronger constant
\[ c_{BSW}=\sup_{a,b}\tau_\M\Big(\exp\big(\ln \Phi^\dagger_A(a)+\ln\Phi^\dagger_B(b)\big)\Big) \pl.\]
which satisfies $c_{BSW}\le c$ by Golden-Thompson inequality.
}
\end{exam}

Another special case is when $\R\subset\A,\B\subset \M$ are sub-algebras with induced traces  $\tau_\A=\tau|_{\A}, \tau_\B=\tau|_{\B}$, and $\tau_\R=\tau|_{\R}$ . Then $E_A$, $E_B$  and $E_R$ are trace preserving conditional expectation.  Petz \cite{petz1991certain} proved that if $E_A(\B)\subset \cR$
then \begin{align}H(E_A(\rho))+H(E_B(\rho))\ge H(\rho)+H(E_\R(\rho))\pl.\end{align}
Theorem \ref{thm:hssa} gives a generalization of the above algebraic SSA inequality
\begin{cor}\label{cor:hssa}
Let $\cR\subset \cA,\cB\subset\cM$ be finite dimensional von Neumann subalgebra with induced traces. Then for any $\rho\in D(\cM)$,
\begin{align}H(E_A(\rho))+H(E_B(\rho))\ge H(\rho)+H(E_\R(\rho))+\log \frac{1}{c}\pl,\end{align}
where the constant $c$ is
\[c=\sup \{\tau_\M(ab) \pl |\pl a\in \A_+, b\in \B_+ \pl \pl,  E_\R(a)=1\pl, \tau_\B(b)=1 \}\pl,\]
In particular, constant $c=1$ if and only if $E_AE_B=E_AE_B=E_R$.
\end{cor}
\begin{proof}The inequality is proved in Theorem \ref{thm:hssa}. Here we discuss the equivalence about $c=1$. Without loss of  generality, we can assume $\tau(1)=1$.
If $c=1$, the for any $b\in D(\B)$, $\norm{E_A(b)}{L_1^\infty(\R\subset\A)} \le 1$. This implies that there exists $\sigma\in D(\R)$ such that
$E_A(b)\le \sigma$.
Note that $\tau(E_A(b) )=\tau(\sigma)=1$. Thus, $E_A(b)=\sigma\in \R$. Hence, we have $E_A(\B)=\cR$, because $\R=E_A(\R)\subset E_A(\B)$. Now we prove $E_B(\A)\subset\cR$. By the definition of $c$, we have for any $a\in \A$,
\[ \tau(ab)=\tau(aE_A(b))=\tau(E_R(a)E_A(b))=\tau(E_R(a)b)\]
Then for any $b\in \B$, $\tau((a-E_R(a)) b)=0$. This implies $E_B(a)=E_B\circ E_R(a)=E_R(a)\in \R$. Therefore, $E_B(\A)=\R$. Finally, by the uniqueness of trace preserving conditional expectation we obtained $E_AE_B=E_R=E_AE_B$.
\end{proof}

\begin{exam}{\rm Recall the Maassan-Uffink uncertainty relation \eqref{eq:MU}: let $H$ be a $d$ dimensional Hilbert space, $\mathcal{X}=\{\ket{x_i}\}_{i=1}^d$ and $\{\ket{z_j}\}_{j=1}^d$ be two orthonormal bases on Hilbert spaces. Consider $\M=\bB(H)$, and $\mathcal{X}, \mathcal{Z}$ are the commutative subalgebra generated by the two basis respectively. The measurement gives the following conditional expectation
\begin{align*}
E_X(\rho)=\sum_{i=1}^d \bra{x_i}\rho \ket{x_i} \ketbra{x_i}\pl, \pl E_Z(\rho)=\sum_{j=1}^d \bra{z_i}\rho \ket{z_i} \ketbra{z_i}
\end{align*}
Berta \emph{et al} \cite{berta2010uncertainty} proved that
\[ H(E_X(\rho))+H(E_Z(\rho))\ge H(\rho)+\log \frac{1}{c}\pl. \]
where  $c=\max_{i,j}|\bra{x_i}z_j\ran|^2=\max_{i,j} \tr(E_X^\dagger( e_i)E_Z^\dagger( e_j) )$. The minimal $c$ can be $\frac{1}{d}$, and in this case $|\bra{x_i}z_j\ran|^2=\frac{1}{d}\pl, \pl \forall \pl 1\le i,j\le d$, for which $\mathcal{X}$ and  $\mathcal{Z}$ are called mutually unbiased bases. In particular, they satisfies commuting square condition
\[ E_XE_Z=E_ZE_X=E_{\mathbb{C}}\pl.\]
}
\end{exam}

\begin{exam}{\rm Consider $\M=\bM_{d^2}$, and  $\A,\B\cong \bM_{d}$ are two subalgebras of $\M=\bM_{d^2}$. If $\M=\A\ten \B$, we have sub-additivity
\[ H(\rho_A)+H(\rho_B)\ge H(\rho_{AB})\pl,\]
where $\rho_A=E_A(\rho),\rho_B=E_B(\rho)$. In general, Corollary \ref{cor:hssa} implies
\[ H(\rho_A)+H(\rho_B)\ge H(\rho)+\log \frac{1}{c}\pl,\pl  c=\sup_{a\in D(\A)\pl, \pl b\in D(\B)}\tr(ab)\pl. \]
Moreover, $c=1$ if and only if $\M=\A\ten \B$. This answer a question of Petz in \cite{petz2007complementarity} .
}
\end{exam}

\section{Uncertainty relation for quantum channels}In this section, we apply Theorem \ref{thm:hssa} to derive the entropic uncertainty relation unde presence of quantum memory. For that, we need to discuss the operator space structure of noncommutative $L_p$-spaces. For simplicity, we consider only  matrix algebras $\bB(H)\cong \bM_n$ equipped with matrix trace $\tr$, whose $L_p$ space is Schatten $p$-class $S_p(H):=S_p^n$. Given a operator space $E$, we define the following norm
\[ \norm{x}{S_p^n(E)}=\inf_{ x= a\cdot y\cdot b} \norm{a}{S_{2p}^n}\norm{y}{\bM_n(E)} \norm{b}{S_{2p}^n} \pl, \pl x\in \bM_n(E)\pl, \]
where the infimum is over all factorization $x=(x_{ij})= (\sum_{k,l}a_{ik}y_{kl}b_{lj})_{ij} , y\in \bM_n(E), a,b\in \bM_n$. This is the vector-valued noncommutative $L_p$-norm introduced by Pisier \cite{pisier1998non}. By  \cite[Lemma 1.7]{pisier1998non}, the completely bounded norm can be characterized by vector-valued noncommutative $L_p$-space. Namely, for any $1\le p\le \infty$,
\begin{align}\norm{T:E\to F}{cb}=\sup_{n} \norm{\id_n \ten T: S_p^n(E)\to S_p^n(F)}{}\pl.\label{eq:pisier}\end{align}
 Here $S_\infty^n(E):=\bM_n(E)$ is the standard operator space structure of $E$. When $E$ is $L_q(\M)$, this is a special case of amalgamated $ L_p $-space,
\[ \pl S_q(K,S_p(H)):= L_q^p( \bB(K)\subset  \bB(K)\ten \bB(H)  ) \pl. \]

Given a density operator $\rho_{MC}\in \bB(H_M\ten H_C) $ on the tensor product Hilbert space $H_M\ten H_C$, the conditional entropy w.r.t $C$ system is defined
\[ H(M|C)_\rho= H(\rho_{MC})-H(\rho_{C})\pl,\]
where $H(\cdot)$  is the von Neumann entropy for matrix trace, $\rho_{C}=\tr_M\ten \id_C (\rho_{MC} )$ is the reduced density operator on $H_C$.

\begin{theorem}\label{thm:ucr}
Let $H_A$, $H_B$ and $H_M$ be finite dimensional Hilbert space. Let $\Phi_A:\bB(H_M)\to \bB(H_A)$ and $\Phi_B:\bB(H_M)\to \bB(H_B) $ be two quantum channels. Then for any Hilbert space $H_C$ and any joint state $\rho_{MC}$ on $H_M\ten H_C$,
\begin{align} H(A|C)_{\Phi_A(\rho)}+H(B|C)_{\Phi_B(\rho)}\ge H(M|C)_\rho+ \log\frac{1}{c} \pl.\label{eq:ucr}\end{align}
where $c$ is the completely bounded norm
\begin{align}\label{eq:c} c=\norm{\Phi_B\circ \Phi_A^\dagger:S_1(H_A)\to \bB(H_B) }{cb}\pl,\end{align}
 \end{theorem}
 \begin{proof}Note that \eqref{eq:ucr} is equivalent to
 \[ H(\Phi_A(\rho))+H(\Phi_B(\rho))\ge H(\rho_{MC})+ H(\rho_C)+\log\frac{1}{c}\]

 Choosing $\M=\bB(H_M\ten H_C), \A=\bB(H_A\ten H_C)\pl, \B=\bB(H_B\ten H_C)$ and  $ \R=\bB(H_C)$ in Theorem \ref{thm:hssa}, we obtain \eqref{eq:ucr} for the constant $c$
 \[ c=\norm{\id_C \ten \Phi_B\circ \Phi^\dagger_A: S_\infty(H_C,S_1(H_A))\to S_\infty (H_C\ten H_B)}{}\]
This yields the completely bounded norm by taking supremum of $H_C$ for all dimensions.
 \end{proof}

 \begin{rem}{\rm It is known \cite{effros1990approximation,blecher1991tensor} that,
 \[ \norm{\Phi_B\circ \Phi_A^\dagger : S_1(H_A)\to \bB(H_B) }{cb}=\norm{C_{\Phi_B\circ \Phi_A^\dagger}}{ \bB(H_A\ten H_B)}\pl, \]
 where
 \[ C_{\Phi_B\circ \Phi_A^\dagger}= \sum_{i,j}e_{ij}\ten \Phi_B\circ \Phi_A^\dagger(e_{ij})\in \bB(H_A\ten H_B)\]
 is the Choi matrix of $\Phi_B\circ \Phi_A^\dagger$. Indeed, by Remark \ref{rem:constant}, we know the constant $c$ can be improved to the state dependent one
 \[  c(\rho)=\norm{\id_C\ten \Phi_B\circ \Phi_A^\dagger\circ\Phi_A(\rho)}{S_1(H_C,\bB(H_B))}\pl.\]}
 \end{rem}

 \begin{exam}{\rm Our result recovers the uncertainty relation of Frank and Lieb \cite{frank2013extended}. Given two positive operator valued measurements $\{E_x\}$ and $\{F_z\}$, define the quantum to classical channel for the measurement
 \[\Phi_A(\rho)=\sum_{x}\tr(\rho E_x)\ketbra{x}\pl,  \Phi_B(\rho)=\sum_{z}\tr(\rho F_z)\ketbra{z},\]
Then
 \[  \Phi_B\circ \Phi_A^\dagger(\rho)=\sum_{x} \tr(E_xF_z)\bra{x}\rho\ket{x} \ketbra{z}\pl, \]
is a classical channel $N(z|x)=\tr(E_xF_z)$ from the commutative system  $\mathbb{C}^X$ to $\mathbb{C}^Z$ with transition matrix as $N(z|x)=\tr(E_xF_z)$. By Smith's lemma \cite{smith1991completely}
\[ c=\norm{\Phi_B\circ \Phi_A^\dagger:\ell_1(X)\to \ell_\infty(Z) }{cb}=\norm{\Phi_B\circ \Phi_A^\dagger: \ell_1(X)\to \ell_\infty(Z) }{}=\max_{x,z}\tr(E_xF_z)\pl,\]
which recovers the maximal overlap of measurement.
 }
 \end{exam}

 \begin{exam}{\rm Consider $\M=\bB(H_A\ten H_B)$, $\A=\bB(H_A)$ and $\B=\bB(H_B)$ with the partial trace channel $\tr_A:\bB(H_A\ten H_B)\to  \bB(H_A)$ and $\tr_B:\bB(H_A\ten H_B)\to  \bB(H_B)$. One have the
 map
 \[ \tr_B\circ (\tr_A)^{\dagger}(X)= \tr_B(X\ten I_B)=I_B\pl, \]
 whose Choi matrix is $\chi=I_A\ten I_B$. Hence, $c=1$ and this recovers the strong sub-additivity
 \[ H(A|C)+H(B|C)\ge H(AB|C)\pl. \]
}
\end{exam}

  Motivated by the examples above, we study the minimum uncertainty under the presence of quantum memory. Let $\Phi_A:\bB(H_M)\to \bB(H_A)$ and $\Phi_B:\bB(H_M)\to \bB(H_B)$ be two quantum channels. For a quantum state $\rho_{MC}\in \bB(H_M\ten H_C)$,
  we define the generalized conditional mutual information
\begin{align}I(\Phi_A,\Phi_B|C)_\rho:= H(A|C)_{\Phi_A\ten \id_C(\rho)}+H(B|C)_{\Phi_B\ten \id_C(\rho)} -H(M|C)_\rho\label{eq:gCMI}\pl.\end{align}
and the minimal uncertainty $\Phi_A$ and $\Phi_B$,
\begin{align}&I(\Phi_A,\Phi_B|C):=\inf_{\rho_{MC}}I(\Phi_A,\Phi_B|C)_\rho\pl ,\\
&I^{sq}(\Phi_A,\Phi_B):=\inf_{H_C}I(\Phi_A,\Phi_B|C)_\rho\pl ,\end{align}
where the infimum runs all density operator $\rho_{MC}\in \bB(H_M\ten H_C)$, and second infimum is over Hilbert space $H_C$ of all dimensions. The notation $I^{sq}$ is motivated by the squashed entanglement introduced  in  \cite{christandl2004squashed}.
Consider the Stinespring dilation of $\Phi_A$ as follows,
 \[\Phi_A(\rho)=\id_A\ten \tr_E(V\rho V^*)\]
where $H_E$ is a Hilbert space, and $V:H_M\to H_A\ten H_E$ is an isometry satisfies $V^*V=1$. As a technical tool we introduce the map
\[\hat{\Phi}_B: \bB(H_A\ten H_E)\to \bB(H_B)\pl, \pl \hat{\Phi}_B(\rho_{AE})=\Phi_B(V^*\rho_{AE} V) \pl.\]
$\hat{\Phi}_B$ is a completely positive and trace non-increasing map, which can be viewed as an extension of $\Phi_B$ by regrading the isometry $V$ as a subspace inclusion. Let $e=VV^*$ be the projection onto the range of $V$. It is clear that $\tr(\hat{\Phi}(\rho))=\tr(\rho)$ if and only if $\rho$ is supported on $e$, i.e. $e\rho e= \rho$. This means the restriction of $\hat{\Phi}_B$ on $\bB(e(H_A\ten H_E))$ is exactly $\Phi_B$, hence trace preserving. We see in the next lemma that the map $\hat{\Phi}_B$ determines $I(\Phi_A,\Phi_B|C)$ and $I^{sq}(\Phi_A,\Phi_B)$.

\begin{lemma}\label{lemma:diff1} Let $1\le p\le \infty$.
Let $H_C$ be a Hilbert space.
Then \begin{align*}&\lim_{p\to 1^+} \frac{\|\id_C\ten \hat{\Phi}_B:
 S_{1}(H_A\ten H_C,S_{p}(H_E))\to S_{1}(H_C,S_{p}(H_B))\|-1}{p-1}=-I(\Phi_A,\Phi_B|C)\\
&\lim_{p\to 1^+} \frac{\|\hat{\Phi}_B:
 S_{1}(H_A,S_{p}(H_E))\to S_{p}(H_B)\|_{cb}-1}{p-1}=-I^{sq}(\Phi_A,\Phi_B)
\end{align*}\end{lemma}
\begin{proof} We define two functions on $[1,\infty]\times \bB(H_{CAE})$,
\begin{align*}&f(p,\rho)=\norm{\id_C\ten\hat{\Phi}_B(\rho)}{S_{1}(H_C,S_{p}(H_B))}\pl,\\
&g(p,\rho)=\norm{\rho}{S_{1}(H_A\ten H_C, S_{p}(H_E))}\pl.
\end{align*}
Denote 
\[ h(p)=\norm{\id_C\ten\hat{\Phi}_B: S_{1}(H_A\ten H_C,S_{p}(H_E))\to S_{1}(H_C,S_{p}(H_B))}{}\pl.\]
Since $\hat{\Phi}_B$ is completely positive, by \cite[Theorem 12]{devetak2006multiplicativity} it suffices to consider its norm for density operators, \begin{align*}
h(p)=\sup_\rho \frac{f(p,\rho)}{g(p,\rho)}\pl.
\end{align*}
Let $p_n\to 1$ be a sequence such that  
\[ \lim_{n\to \infty }\frac{h(p_n)-1}{p_n-1}=\limsup_{p\to 1^+}\frac{h(p)-1}{p-1}  \pl.\] 
Suppose $\rho_n$ is a sequence such that attains $h(p_n)$ for each $p_n$. Without loss of generality, we can assume $\rho_n\to \rho$  converges.
Then
\begin{align*}
  \limsup_{p\to 1^+}\frac{h(p)-1}{p-1}=&\lim_{n\to \infty }\frac{h(p_n)-1}{p_n-1} =\lim_{n\to \infty }\frac{f(p_n,\rho_n)-1}{p_n-1}\\
  =&\lim_{n\to \infty } \frac{1}{g(p_n,\rho_n)}(
  \frac{f(p_n,\rho_n)-1}{p-1} - \frac{g(p_n,\rho_n)-1}{p-1})
  \pl. \end{align*}
Note that we should have \[\lim_{n\to \infty}f(p_n,\rho_n)=f(1,\rho)=1,\] otherwise the above limit equals $-\infty$. Note that by complex interpolation, $h(p)\le h(1)^{\frac{1}{p}} h(\infty)^{(1-\frac{1}{p}) }=h(\infty)^{(1-\frac{1}{p}) }$ and
\[  \limsup_{p\to 1^+}\frac{h(p)-1}{p-1}\le \limsup_{p\to 1^+}\frac{h(\infty)^{(1-\frac{1}{p}) }-1}{p-1}=\ln h(\infty)<\infty\pl,\]
which leads to a contradiction. Thus we have $\tr(\id_C\ten \hat{\Phi}_B(\rho_1))=1$, which means $\rho$ is supported on $e H_{AE}\cong H_M$.
By Lemma \ref{lemma:diff}
\begin{align*}\lim_{n\to \infty}\frac{f(p_n,\rho_{p_n})-1}{p_n-1} =&\lim_{n\to \infty}\frac{\norm{\id_C\ten\hat{\Phi}_B(\rho_n)}{S_{1}(H_C,S_{p_n}(H_B))}-1}{p_n-1} = H(\id_C\ten\hat{\Phi}_B(\rho) )-H(\rho_{C} )\\
\lim_{n\to \infty}\frac{g(p_n,\rho_{p_n})-1}{p_n-1} =&\lim_{n\to \infty}\frac{\norm{\rho_n}{S_{1}(H_A\ten H_C,S_{p_n}(H_E))}-1}{p_n-1}
= H(\rho_{CM} )-H(\Phi_A(\rho_{MC}))\\ =&H(\rho_{CM} )-H(\rho_{AC})
\end{align*}
 Therefore,
  \begin{align*}
  \limsup_{p\to 1^+}\frac{h(p)-1}{p-1}=& -H(\id_C\ten\hat{\Phi}_B(\rho) )+H(\rho_{C} ) +H(\rho_{CM} )-H(\Phi_A(\rho_{MC}))\\
  =& -H(A|C)_{\Phi_A(\rho)}-H(B|C)_{\Phi_B(\rho)}+H(M|C)_\rho=-I(\Phi_A,\Phi_B|C)_\rho\\
  \le  & -\inf_\rho I(\Phi_A,\Phi_B|C)_\rho= -I(\Phi_A,\Phi_B|C)
  \pl. \end{align*}
For the other direction, we assume that $I(\Phi_A,\Phi_B|C)$ is attained by $\omega_{MC}$. Then 
\begin{align*}
-I(\Phi_A,\Phi_B|C)=&-I(\Phi_A,\Phi_B|C)_\omega\\ =& -H(\id_C\ten\hat{\Phi}_B(\omega) )+H(\omega_{C} ) +H(\omega_{CM} )-H(\Phi_A(\omega_{MC}))\\
=&\lim_{p\to 1^+} \frac{\frac{f(p, \omega)}{g(p,\omega)}-1}{p-1} \\ \le &\liminf_{p\to 1^+} \frac{\sup_{\rho}\frac{f(p, \omega)}{g(p,\omega)}-1}{p-1}\\ =& \liminf_{p\to 1^+} \frac{h(p)-1}{p-1}\pl.
\end{align*}
The second asserted equality follows from taking supremum over all $H_C$. \end{proof}

In the following, we use the short notation $H_{AB}:=H_A\ten H_B$.
\begin{lemma}\label{lemma:add}
Let $\hat{\Phi}_j: \bB(H_{A_jE_j})\to \bB(H_{B_j}), j=1,2$ be two linear maps respectively. Then
\begin{align}
&\norm{\hat{\Phi}_1\ten \hat{\Phi}_2: S_{1}(H_{A_1A_2},S_{p}(H_{E_1E_2}))\to S_{p}(H_{B_1B_2})}{cb}\nonumber\\=&\norm{\hat{\Phi}_1:S_{1}(H_{A_1},S_{p}(H_{E_1}))\to S_{p}(H_{B_1}) }{cb}\norm{\hat{\Phi}_2:S_{1}(H_{A_2},S_{p}(H_{E_2}))\to S_{p}(H_{B_2})}{cb}\pl.\label{eq:eq}
\end{align}
\end{lemma}
\begin{proof}We will repeatedly use the noncommutative version of the Minkowski's inequality \cite[Corollary 1.10]{pisier1998non} that for any operator space $E$, the identity map
\begin{equation}\label{ps3}
  \id: S_p(H_A;S_q(H_B;E)) \to S_q(H_B;S_p(H_A;E))
\end{equation}
is a complete contraction provided that $q\gl p$. We write
\[ \hat{\Phi}_1\ten \hat{\Phi}_2: S_{1}(H_{A_1A_2},S_{p}(H_{E_1E_2}))\to S_{p}(H_{B_1B_2})\]
as the composition of the following four maps,
\begin{align*}
& S_{1}(H_{A_1A_2},S_{p}(H_{E_1E_2})) \\ \overset{\id}{\longrightarrow}&S_{1}(H_{A_1},S_{p}(H_{E_1},S_{1}(H_{A_2},S_{p}(H_{E_2}))))\\
\overset{\id\ten \hat{\Phi}_2}{\longrightarrow} &S_{1}(H_{A_1},S_{p}(H_{E_1},S_{p}(H_{B_2})))\\
\overset{\id}{\longrightarrow} &S_{p}(H_{B_2},S_{1}(H_{A_1},S_{p}(H_{E_1})))\\
\overset{\id\ten \hat{\Phi}_1}{\longrightarrow} &S_{p}(H_{B_1B_2})
\end{align*}
The first map and third map are complete contractions by \eqref{ps3}. Let us recall the Pisier lemma \ref{eq:pisier} that for any linear map $T:E\to F$ and $1\le p,q\le \infty$
 \begin{align*}
  &\|\id_H\ten T:S_p(H,E)\to S_p(H,F)\|_{}
  \lel  \|\id_H\ten T:S_q(H,E)\to S_q(H,F)\|_{}\pl .
  \end{align*}
Applying this property twice, we have for the second  map
\begin{align*} &\norm{\id_{A_1E_1}\ten \Phi_2 :  S_{1}(H_{A_1},S_{p}(H_{E_1},S_{1}(H_{A_2},S_{p}(H_{E_2}))))\to S_{1}(H_{A_1},S_{p}(H_{E_1},S_{p}(H_{B_2})))}{cb}\\ \le &\norm{\Phi_2: S_{1}(H_{A_2},S_{p}(H_{E_2}))\to S_{p}(H_{B_2}))}{}\end{align*}
and the fourth map
\begin{align*}
 &\norm{\id_{B_2}\ten \Phi_1 :  S_{p}(H_{B_2},S_{1}(H_{A_1},S_{p}(H_{E_1})))\to S_{p}(H_{B_1B_2})}{cb} \\ \le &\norm{\Phi_1: S_{1}(H_{A_1},S_{p}(H_{E_1}))\to S_{p}(H_{B_1}))}{}
\end{align*}
Thus, we show the ``$\le$'' direction in the desired equality \eqref{eq:eq}. The other direction follows from tensor product elements. 
\end{proof}
We obtain the following additivity result.
\begin{theorem} 
$I^{sq}$ is additive. That is, for two pairs of quantum channels $(\Phi_A,\Phi_B)$ and $(\Psi_A,\Psi_B)$,
\[I^{sq}(\Phi_A\ten \Psi_A,\Phi_B\ten\Psi_B)=I^{sq}(\Phi_A,\Phi_B)+I^{sq}(\Psi_A,\Psi_B)\pl.\]
\end{theorem}
\begin{proof} By Lemma \ref{lemma:diff1} and  Lemma \ref{lemma:add},
\begin{align*}
&-I^{sq}(\Phi_A\ten \Psi_A,\Phi_B\ten\Psi_B)\\ =&\lim_{p\to 1^+}\frac{\norm{\hat{\Phi}_B\ten \hat{\Psi}_B: S_{1}(A_1A_2,S_{p}(E_1E_2))\to S_{p}(B_1B_2)}{cb}-1}{p-1}\\
=&\lim_{p\to 1^+}\frac{\norm{\hat{\Phi}_B: S_{1}(A_1,S_{p}(E_1))\to S_{p}(B_1)}{cb}\norm{\hat{\Psi}_B: S_{1}(A_2,S_{p}(E_2))\to S_{p}(B_2)}{cb}-1}{p-1}\\
=&\lim_{p\to 1^+}\norm{\hat{\Psi}_B: S_{1}(A_2,S_{p}(E_2))\to S_{p}(B_2)}{cb}\frac{\norm{\hat{\Phi}_B: S_{1}(A_1,S_{p}(E_1))\to S_{p}(B_1)}{cb}-1}{p-1} \\ &+ \lim_{p\to 1^+}\frac{\norm{\hat{\Psi}_B: S_{1}(A_2,S_{p}(E_2))\to S_{p}(B_2)}{cb}-1}{p-1}\\
=&-I^{sq}(\Phi_A,\Phi_B)- I^{sq}(\Psi_A,\Psi_B) \qedhere
\end{align*}
\end{proof}
\begin{rem}{\rm The above additivity results can be extended to minimal uncertainty with parameters
\begin{align*}
I^{sq}_\al(\Phi_A,\Phi_B|C):=\inf_{\rho^{MC}}\al_A H(A|C)+\al_B H(B|C) -\al_M H(M|C)\pl ,
\end{align*}
where $\al=(\al_A,\al_B,\al_M)$ are non-negative parameters satisfying $0\le \al_A\le \al_M\le \al_B$. Indeed, similar to Example \ref{exam:s1p} and Lemma \ref{lemma:diff1}, we have
\begin{align*} &\lim_{p\to 1^+}\frac{1}{p-1}(\|\rho^{CA}\|_{S_{q_1}(H_C,S_{q_2}(H_A))}-1) \lel (\al_2-\al_1)H(C)-\al_2H(CA)\pl, \\
&\lim_{p\to 1^+}\frac{1}{p-1}(\norm{id_C\ten \hat{\Phi}_B :S_{q_{1}}(H_{CA},S_{q_2}(H_E))\to S_{q_1}(H_C,S_{q_3}(H_B)) }{}-1)=-I_\al(\Phi_A,\Phi_B|C)
\end{align*}
where $q_1, q_2$ $q_3$ are functions of $p$ satisfying following relations \begin{align*}1-\frac{1}{q_j(p)}=\al_j(1-\frac{1}{p})\pl, \pl j=1,2,3.
 \end{align*}
 The additivity of $I^{sq}_\al(\Phi_A,\Phi_B|C)$ follows similarly via the multiplicativity of CB-norm in Lemma \ref{lemma:add}. The reader are referred to \cite{gao2018uncertainty} for the details.
}\end{rem}

\section{Strong sub-additivity of relative entropy}
In this section, we discuss a generalized strong sub-additivity for relative entropy. Our motivation is the following result of Petz. Recall that for two density operators $\rho\in D(\M)$ and $\sigma\in D_+(\M)$, the relative entropy is
\[ D(\rho||\sigma)=\tr(\rho\log \rho -\rho\log \sigma) \pl.\]

\begin{theorem}[Petz \cite{petz1991certain}]\label{pssa} Let $\M$ be a $C^*$-algebra, and $\A,\B\subset \M$ be a subalgebra. Let  $\si$ be a faithful state of $\M$ and assume that there is a $\sigma$-preserving conditional expectation $E_A^\dagger:\M\to \A$. If $E_A^\dagger(\B)=\R$ is a subalgebra, the for any state $\rho$,
\[D(\rho||\si)+D(\rho_R||\si_R)\ge D(\rho_A||\si_A)+D(\rho_B||\si_B)\pl,\]
where $\rho_A=\rho|_{\A}, \sigma_A=\sigma|_{\A}$ are the restriction state on $\A$ and similarly for subalgebra $\cB$ and $\cR$.
\end{theorem}
We now present a quantitative extension of above theorem.
\begin{theorem}\label{thm:rssa}
Let $\cA,\cB$ and $\cM$ be finite dimensional von Neumann algebras equipped with trace $\tau_\A,\tau_\B$ and $\tau_\M$. Let $\Phi_A: \M\to \A$ and $\Phi_B: \M\to \B$ be two quantum channels. Suppose $\R \subset\B$ is a subalgebra, and assume that $\sigma\in D_+(\M)$ is a density operator such that there exists a conditional expectation $E_R^\dagger:\cB\to \cR$ preserving $\Phi_B(\sigma)$. Then for any $\rho\in D(\M)$, we have
\[D(\rho||\si)+D(E_R\circ\Phi_B(\rho)||\Phi_B(\si))\ge D(\Phi_A(\rho)||\Phi_A(\si))+D(\Phi_B(\rho)||\Phi_B(\si))-\kappa \]
Here, the constant $\kappa$ is
\begin{align*}
\kappa=&\int_{\bR} \al(t)\log c(t) dt\pl, \pl \al(t)=\frac{\pi}{2(\cosh(\pi t)+1)}\pl ,\\
c(t)=&\sup_{b} \tau_\M\Big(\Phi_B^\dag(b)
\Phi_\A^\dag\big(\Phi_\A(\rho)^{\frac{1+it}{2}}\Phi_A(\si)^{\frac{-1-it}{2}}
\big)\si\Phi_\A^\dag\big(\Phi_\A(\rho)^{\frac{1+it}{2}}\Phi_A(\si)^{\frac{-1-it}{2}}
\big)^*\Big)\pl,
\end{align*}
where the supremum is for all $b\in \B_+$ such that $E_R^\dagger(b)=1$.
\end{theorem}

The proof is divided into two steps.
Given $\sigma\in D_+(\M)$ and  $\rho\in D(\M)$, we define the parameter
\[\la(p)= \norm{\si^{-\frac{1}{2}}\rho\si^{-\frac{1}{2}}}{p,\si}=\norm{\si^{-\frac{1}{2p'}}\rho\si^{-\frac{1}{2p'}}}{p}\]
For $1\le p\le \infty$, we denote
\[\rho_A=\Phi_A(\rho)\pl, \rho_B=\Phi_B(\rho)\pl,  \rho_R=E_R\circ \Phi_B(\rho)\pl, \]
and similarly for $\sigma_A$, $\sigma_B$ and  $\sigma_R$. Recall that the condition expectation $E_R^\dagger:\cB\to \cR$ induce a natural weight $\sigma_{\tr}=E_R(1)\in \R'\subset \B$.
\begin{lemma}\label{lemma:dif}For $p>1$,
define
\begin{align*}
\Delta(p):=\la(p)^{-1}\norm{\si_B^{-\frac{1}{2}} \Phi_B\Big(\Phi_A^\dag\big(\rho_A^{\frac{1}{2p'}}\sigma_A^{-\frac{1}{2p'}}\big)\rho\Phi_A^\dag\big(\sigma^{-\frac{1}{2p'}}\rho_A^{\frac{1}{2p'}}\big)\Big)\si_B^{-\frac{1}{2}}}{L_1^p(\R\subset \B, \sigma_\tr)}
\end{align*}
We have
\[\lim_{p\to 1^+} \frac{\Delta_p(p) -1}{p-1}\ge D(\rho_A||\sigma_A)+D(\rho_B||\si_B)- D(\rho_R||\sigma_R)-D(\rho||\si)\pl.\]
\end{lemma}
\begin{proof}First, $\la_1=1$ and by Lemma \ref{lemma:dp}
\[\lim_{p\to 1^+} \frac{\la(p)^{-1} -1}{p-1}= \lim_{p\to 1^+}\frac{\norm{\si^{-\frac{1}{2p'}}\rho\si^{-\frac{1}{2p'}}}{p}^{-1}-1}{p-1} =-D(\rho||\si ) \pl. \]
Define
\[x_p=\Phi_A^\dag\big(\rho_A^{\frac{1}{2p'}}\sigma_A^{-\frac{1}{2p'}}\big)\rho\Phi_A^\dag\big(\sigma_A^{-\frac{1}{2p'}}\rho_A^{\frac{1}{2p'}}\big)\pl.\]
Denote $s(\rho)$ as the support of $\rho$. When $p\to 1,\frac{1}{p'}=\frac{p-1}{p}\to 0$, we have
\begin{align*}
\lim_{p\to 1^+}x_p=\Phi_A^\dag(s(\rho_A))\rho\Phi_A^\dag(s(\rho_A))=\rho
\end{align*}
In fact, for any positive $0\le y\le 1$, $\Phi_A (\rho^{\frac{1}{2}}y\rho^{\frac{1}{2}})\le \Phi_A (\rho)=\rho_A$, so $s(\Phi_A (\rho^{\frac{1}{2}}y\rho^{\frac{1}{2}}))\le s(\rho_A)$. Hence,
\begin{align*} \tau_\M(y\rho^{\frac{1}{2}}\Phi_A^\dag(s(\rho_A))\rho^{\frac{1}{2}})
=&\tau_\A\Big(\Phi_A (\rho^{\frac{1}{2}}y\rho^{\frac{1}{2}})s(\rho_A)\Big)
=\tau_\A\Big(\Phi_A (\rho^{\frac{1}{2}}y\rho^{\frac{1}{2}})\Big)
=\tau_\M\Big(\rho y\Big)\pl.
\end{align*}
Therefore,
\[\rho^{\frac12}\Phi_A^\dagger(s(\rho_A))\rho^{\frac12}=\rho\pl , \pl  \Phi_A^\dagger(s(\rho_A)) \rho \Phi_A^\dagger(s(\rho_A))=\rho \pl. \]
We split the desired limit as the following three parts
\begin{align*}\lim_{p\to 1^+}\frac{\Delta_p(p)-1}{p-1}=& \lim_{p\to 1^+}\norm{\si_B^{-\frac{1}{2}} \Phi_B(x_p)\si_B^{-\frac{1}{2}} }{L_1^p(\R\subset \B, \sigma_\tr)} \frac{\la(p)^{-1}-1}{p-1}\\ &+
\frac{\norm{\si_B^{-\frac{1}{2}} \Phi_B(x_p)\si_B^{-\frac{1}{2}} }{L_1^p(\R\subset \B, \sigma_\tr)}-\tau_\B(\Phi_B(x_p))}{p-1}+ \frac{\tau_\M(x_p)-1}{p-1}
\\ :=&\text{I}+\text{II}+\text{III}
\end{align*}
By $\norm{\si_B^{-\frac{1}{2}} \Phi_B(x_p)\si_B^{-\frac{1}{2}} }{L_1^p(\R\subset \B, \sigma_\tr)}\to 1$, the first part is calculated.
The limits for part II and III are as follows,
\begin{align*}
&\lim_{p \to 1^+} \text{II}(p)\ge D(\rho_B||\si_\tr)-D(E_R(\rho_B)||\si_\tr)=D(\rho_B||\sigma_B)-D(E_R(\rho_B)||\sigma_B)\pl,\\
& \lim_{p \to 1^+} \text{III}(p)\ge D(\rho_A||\si_A)
\end{align*}
The part II follows from Lemma \ref{lemma:dp1} and $\displaystyle\lim_{p\to 1^+}\Phi_B(x_p)=\Phi_B(\rho)$. For part III, note that for a positive $a$
\[ \lim_{q\to 0}a^q=s(a)\pl, \pl \left.\frac{d}{dq} a^q \right\vert_{q=0}= s(a)\log a\pl  \pl.\]
Because $s(\rho_A)\le s(\sigma_A)$, we have
\begin{align*}
\lim_{p\to 1^+} \text{III}(p)=&\lim_{p\to 1}\frac{\tau_\M(x_p)-1}{p-1}
\\=&-\frac{1}{2}\tau_\M\Big(\Phi_A^\dag\big( s(\rho_A)\log (\sigma_A)\big)\rho \Phi_A^\dag\big(s(\rho_A)\big)\Big)
+\frac{1}{2}\tau_\M\Big(\Phi_A^\dag\big(\log\rho_A \big)\rho \Phi_A^\dag\big(\rho_A\big)\Big) \\
&+\frac{1}{2}\tau_\M\Big(\Phi_A^\dag\big( s(\rho_A)\big)\rho \Phi_A^\dag\big(\log \rho_A\big)\Big)
-\frac{1}{2}\tau_\M\Big(\Phi_A^\dag\big( s(\rho_A)\big)\rho \Phi_A^\dag\big(\log \sigma_A s(\rho_A)\big)\Big)
\\=&-\tau_\M\Big(\rho_A\log (\sigma_A) s(\rho_A)\Big) +\tau_\M\Big(\rho_A\log (\rho_A)s(\rho_A)\Big)
\\=&D(\rho_A||\si_A)\pl.
\end{align*}
Combining the three parts above, we finish the proof.
\end{proof}

Fix $1<p< \infty$, define the analytic family of operator
\[\rho: \{0\le \Re(z)\le 1\}\to \M \pl ,\pl \rho(z)= \la_p^{-pz} \si^{-\frac{z}{2}} |\si^{-\frac{1}{2p'}}\rho\si^{-\frac{1}{2p'}}|^{pz}\si^{-\frac{z}{2}}\]
Note that \[\rho(\frac{1} {p})=\frac{\si^{\frac{-1}{2}}\rho\si^{\frac{-1}{2}}}{\norm{\si^{-\frac{1}{2}}\rho\si^{-\frac{1}{2}}}{p,\si}}=\la_p^{-1}\si^{\frac{-1}{2}}\rho\si^{\frac{-1}{2}} ,\]
and
\begin{align*}&\norm{\rho(it)}{\infty}=\norm{\si^{\frac{-it}{2}} |\si^{-\frac{1}{2p'}}\rho\si^{-\frac{1}{2p'}}|^{ipt}\si^{\frac{-it}{2}}}{\infty}\le 1\\
&\norm{\rho(1+it)}{1,\sigma_\tr}=\la_p^{-p}\norm{\si^{\frac{-1-it}{2}} |\si^{-\frac{1}{2p'}}\rho\si^{-\frac{1}{2p'}}|^{p+ipt}\si^{\frac{-1-it}{2}}}{1,\si}\le 1\pl.
\end{align*}
For $\Delta(\rho,\gamma)$, we have the following estimate:
\begin{lemma}\label{lemma:pest}
For any $1<p<\infty$ and $\gamma\in \cR_+$,
\[\lim_{p\le 1^+}\frac{\Delta(p)-1}{p-1}\le \kappa\pl,\]
where
\begin{align*}
\kappa=&\int_{\bR}  \al(t) \log c(t) dt\pl , \pl  \al(t) =\frac{\pi}{2(\cosh(\pi t)+1)} \\
c(t)=&\sup_{t\in \mathbb{R}}\norm{ \si_\tr^{-\frac{1}{2}}
\Phi_B\Big(\Phi_A^\dag\big(\rho_A^{\frac{1+it}{2}}\sigma_A^{\frac{-1-it}{2}}
\big)\si\Phi_A^\dag\big(\sigma_A^{\frac{-1+it}{2}}\rho_A^{\frac{1-it}{2}}
\big)
\Big)
\si_\tr^{-\frac{1}{2}}}{L_1^\infty(\R\subset \B, \sigma_\tr)}\\
=&\sup_{b}\tau_\M(\Phi_B^\dagger(b)\Phi_A^\dag\big(\rho_A^{\frac{1+it}{2}}\sigma_A^{\frac{-1-it}{2}}
\big)\si\Phi_A^\dag\big(\sigma_A^{\frac{-1+it}{2}}\rho_A^{\frac{1-it}{2}}
\big) )
\end{align*}
where the supremum is for all $b\in \B_+$ such that $E_R^\dagger(b)\le 1$.
\end{lemma}
\begin{proof}
Fix  $1<p <\infty$, we consider the following analytic family of operators,
\[A(z)=\si_B^{-\frac{1}{2}}
\Phi_B\Big(\Phi_A^\dag\big(
\rho_A^{\frac{1-z}{2}}\sigma_A^{\frac{z-1}{2}}\big)\rho (z)  \Phi_A^\dag\big(\sigma_A^{\frac{z-1}{2}}
\rho_A^{\frac{1-z}{2}}\big)\Big)
\si_B^{-\frac{1}{2}}\pl.\]
Note that
\[ \norm{A(\frac{1}{p})}{L_1^p(\R\subset \B,\sigma_\tr)}=\Delta(p)\]
For $z=1+it$,\begin{align*}\norm{A(1+it)}{1,\sigma_{\tr}}&=\norm{\Phi_B\Big(\Phi_A^\dag\big(
\rho_A^{\frac{it}{2}}\sigma_A^{\frac{-it}{2}}\big)\si^\frac{1}{2}\rho (1-it) \si^\frac{1}{2} \Phi_A^\dag\big(\sigma_A^{\frac{-it}{2}}
\rho_A^{\frac{it}{2}}\big)\Big)}{1}
\\ &\le \norm{\Phi_A^\dag\big(
\rho_A^{\frac{it}{2}}\sigma_A^{\frac{-it}{2}}\big)\si^\frac{1}{2}\rho (1-it) \si^\frac{1}{2} \Phi_A^\dag\big(\sigma_A^{\frac{-it}{2}}
\rho_A^{\frac{it}{2}}\big)}{1}
\\ &\le \norm{\si^\frac{1}{2}\rho (1-it) \si^\frac{1}{2} }{1}
\\ &\le \la_p^{-p}\norm{\si^{it}|\si^{-\frac{1}{2p'}}\rho \si^{-\frac{1}{2p'}}|^{p-ipt}\si^{-it}  }{1}=1
\end{align*}
For $z=it$,
\begin{align*}\norm{A(it)}{L_1^\infty(\R\subset \B, \sigma_\tr)}=&\norm{\si_B^{-\frac{1}{2}}
\Phi_B\Big(\Phi_A^\dag\big(\rho_A^{\frac{1+it}{2}}\sigma_A^{\frac{-1-it}{2}}
\big)\si^{\frac{1}{2}}\rho (it)\si^{\frac{1}{2}}\Phi_A^\dag\big(\sigma_A^{\frac{-1-it}{2}}\rho_A^{\frac{1+it}{2}}
\big)
\Big)
\si_B^{-\frac{1}{2}}}{L_1^\infty(\R\subset \B, \sigma_\tr)}
\end{align*}Let $\gamma_1,\gamma_2\in \cR_+$ be two arbitrary positive elements in $\cR$ with $\norm{\gamma}{1,\sigma_\tr}=1$.
Denote \[X_1=\gamma_1^{-\frac{1}{2}}\si_B^{-\frac{1}{2}},\  X_1=\gamma_2^{-\frac{1}{2}}\si_B^{-\frac{1}{2}},\  Y(t)=\Phi_A^\dag(\rho_A^{\frac{1+it}{2}}\sigma_A^{\frac{-1-it}{2}})\pl. \] We have
\begin{align*} \norm{A(it)}{L_1^\infty(\R\subset \B, \sigma_\tr)}=&\norm{\si_B^{-\frac{1}{2}}\Phi_B(Y(t) \sigma^{1/2} \rho(it) \sigma^{1/2}Y(-t)^* )\si_B^{-\frac{1}{2}}}{L_1^\infty(\R\subset \B, \sigma_\tr)}
\\ =&\inf_{\gamma_1,\gamma_2} \norm{ \gamma_1^{-\frac{1}{2}}\si_B^{-\frac{1}{2}}\Phi_B(Y(t) \sigma^{1/2} \rho(it) \sigma^{1/2}Y(-t)^* )\si_B^{-\frac{1}{2}}\gamma_2^{-\frac{1}{2}}}{\infty}\end{align*}
Note that
\begin{align}
&\left[\begin{array}{cc}
X_1\Phi_B(Y(t) \sigma Y(t)^* )X_1^*&
A(it)\\
A(-it)& X_2\Phi_B(Y(-t) \sigma Y(-t)^* )X_2^*
\end{array}
\right]\nonumber \\ =&\left[\begin{array}{cc}
X_1 &0 \\ 0&X_2
\end{array}\right] \cdot \nonumber\\ &
\Phi_B \left( \left[\begin{array}{cc}
Y(t) &0 \\ 0& Y(-t)
\end{array}\right]
\left[\begin{array}{c}
 \sigma^{1/2}\rho(it)\\
\sigma^{1/2}
\end{array} \right]\left[\begin{array}{c}
 \sigma^{1/2}\rho(it)\\
\sigma^{1/2}
\end{array} \right]^*
\left[\begin{array}{cc}
Y(t)^* &0 \\ 0& Y(-t)^*
\end{array}\right]
\right)\nonumber \\ & \cdot
\left[\begin{array}{cc}
X_1^* &0 \\ 0&X_2^*
\end{array}\right] \nonumber\\ \ge &0 \label{eq:order}
\end{align}
Denote
\begin{align*}c(t):=&\norm{\Phi_B(Y(t) \sigma Y(t)^* )}{L_1^\infty(\R\subset \B, \sigma_\tr)}\\ =& \inf_{\gamma}\norm{ \gamma^{-\frac{1}{2}} \si_B^{\frac{-1}{2}}\gamma^{\frac{-1-it}{2}}
\Phi_B\Big(\Phi_A^\dag\big(\rho_A^{\frac{1+it}{2}}\sigma_A^{\frac{-1-it}{2}}
\big)\si\Phi_A^\dag\big(\sigma_A^{\frac{-1+it}{2}}\rho_A^{\frac{1-it}{2}}
\big)
\Big)
\gamma^{\frac{-1+it}{2}}\si_B^{\frac{-1}{2}}\gamma^{-\frac{1}{2}} }{\infty} \end{align*}
Then by \eqref{eq:order} we have
\[ \norm{A(it)}{L_1^\infty(\R\subset \B, \sigma_\tr)}\le \sqrt{c(t )c(-t)}\]
Now, by Hirschma interpolation theorem \cite{hirschman1952convexity} (see also \cite[Lemma 3.2]{junge2018universal}), we have
\begin{align*}\log \norm{A(\frac{1}{p})}{L_1^p(\R\subset \B, \sigma_\tr)}\le &\int_{\bR} \beta_{\frac{1}{p}}(t)\log \norm{A(it)}{1,\sigma_{\tr}}^{\frac{1}{p}} + \al_{\frac{1}{p}}(t)\log \norm{A(1+it)}{L_1^\infty(\R\subset \B, \sigma_\tr)}^{1-\frac{1}{p}}dt
\\ \le & \frac{p-1}{p}  \int_{\bR}\frac{1}{2} \al_{\frac{1}{p}}(t)(\log c(t)+\log c(-t))dt
\\ =&\frac{p-1}{p}  \int_{\bR}\al_{\frac{1}{p}}(t)\log c(t)dt
\end{align*}
where
\[\al_\frac{1}{p}(t)=\frac{\sin(\frac{\pi}{p})}{2(1-\frac{1}{p})(\cosh(\pi t)-\cos(\pi \theta)) }\pl,\]
and
\[ \lim_{p\to 1^+}\al_\frac{1}{p}(t)=\frac{\pi}{2(\cosh(\pi t)+1)}:=\al(t)\pl.\]
Hence, we have
\begin{align*}
\lim_{p\to 1^+}\frac{\Delta(p)-1}{p-1}=&\lim_{p\to 1^+}\frac{\Delta(p)^p-1}{p-1}
\\ =& \lim_{p\to 1^+}\frac{p\log\Delta(p)}{p-1}
\\ =& \lim_{p\to 1^+}\frac{p\log\norm{A(\frac{1}{p})}{L_1^p(\R\subset \B, \sigma_\tr)}}{p-1}
\\ \le& \lim_{p\to 1^+}
\int_{\bR}\al_{\frac{1}{p}}(t)\log c(t)dt= \int_{\bR}\al(t)\log c(t)dt.
\end{align*}
This finishes the proof
\end{proof}

Theorem \ref{thm:rssa} now follows from Lemma \ref{lemma:dif} and Lemma \ref{lemma:pest}. We discuss some special cases.
\begin{exam}{\rm If  $\A, \R=\mathbb{C}1$ are trivial subalgebras, we the obtain data processing inequality
\[ D(\rho||\si)\ge D(\Phi_B(\rho)||\Phi_B(\si))
\]
as the constant are
\begin{align*}
&c(t)=\sup_{\norm{\pl b\pl }{\si_B,1}=1}\tau_M(\Phi_B^\dag(b)\si
)=\tau_B(b\Phi_B(\si)
)=1\\
&\kappa=0
\end{align*}
}
\end{exam}

\begin{exam}{\rm If $\B=\R=\mathbb{C}$ are trivial subalgebra, we have
\[ D(\rho||\si)\ge D(\Phi_A(\rho)||\Phi_A(\si))-\kappa\pl.\]
and $\kappa\le 0$. Because
\begin{align*}
c(t)=&\tau_M\Big(
\Phi_A^\dag\big(\Phi_A(\rho)^{\frac{1+it}{2}}\Phi_A(\si)^{\frac{-1-it}{2}}
\big)\si\Phi_A^\dag\big(\Phi_A(\rho)^{\frac{1+it}{2}}\Phi_A(\si)^{\frac{-1-it}{2}}
\big)^*\Big)
\\ \le &\tau_M\Big(
\Phi_A^\dag\big( \Phi_A(\si)^{\frac{-1+it}{2}}\Phi_A(\rho)^{\frac{1-it}{2}}\Phi_A(\rho)^{\frac{1+it}{2}}\Phi_A(\si)^{\frac{-1-it}{2}}
\big)\si\Big)
\\ =&\tau_A\Big(
\Phi_A(\si)^{\frac{-1+it}{2}}\Phi_A(\rho)\Phi_A(\si)^{\frac{-1-it}{2}}
\Phi_A(\si)\Big)
\\ =&\tau_A(\Phi_A(\rho))\\ =& 1\pl,
\end{align*}
Here, we used Kadison-Schwarz inequality  $\Phi^\dagger(x^*)\Phi^\dagger(x)\le \Phi^\dagger(x^*x)$.
This gives an improvement for data processing inequality. Moreover, our constant $\kappa$ is tight in the following sense: if $\kappa=0$, because  $\al(t)dt$ is a probability measure, we have
\[ c(t)=1\pl , \pl \forall\pl t\in \bR\pl.\]
This means
\[ \Phi_A^\dag\big(\Phi_A(\rho)^{\frac{1+it}{2}}\Phi_A(\si)^{\frac{-1-it}{2}}
\big)^*\Phi_A^\dag\big(\Phi_A(\rho)^{\frac{1+it}{2}}\Phi_A(\si)^{\frac{-1-it}{2}}
\big)=\Phi_A^\dag\big(\Phi_A(\si)^{\frac{-1-it}{2}}\Phi_A(\rho)\Phi_A(\si)^{\frac{-1-it}{2}}
\big)\pl.\]
Hence for all $t \in \bR$, $\Phi_A(\rho)^{\frac{1+it}{2}}\Phi_A(\si)^{\frac{-1-it}{2}}$ is in the multiplicative domain  of $\Phi_A^\dagger$, which further extends to $\{\Phi_A(\rho)^ {z}\Phi_A(\si)^{-z}, z\in \mathbb{C}\}$ by analytic extension. Note that, this condition is equivalent to
\[ D(\rho||\si)= D(\Phi_A(\rho)||\Phi_A(\si))\]
and there exists a channel  $\Psi$ such that $\Psi \circ \Phi_A(\rho)=\rho$ and  $\Psi \circ \Phi_A(\sigma)=\sigma$ (see \cite{petz2003monotonicity}).
Therefore, we have
\[ \kappa =0 \Longleftrightarrow D(\rho||\si)= D(\Phi_A(\rho)||\Phi_A(\si))\pl. \]
}
\end{exam}

\begin{exam}{\rm
Let $ \A,\B\subset \M$ be subalgebras and $\Phi_A=E_A,\Phi_B=E_B$ be the adjoint map of the inclusions.  We have
\[ D(\rho||\si)+D(E_R\circ E_B(\rho)||E_B(\sigma))\ge D(E_A(\rho)||E_A(\si))+D(E_B(\rho)||\si_B)-\int_{\mathbb{R}} \al(t)\log c(t)dt\pl.\]
Here the constant is
\begin{align*}
c(t)&=\sup_{b\in \cB, E_R(b)=1}\tau_M\Big(b
E_A(\rho)^{\frac{1+it}{2}}E_A(\si)^{\frac{-1-it}{2}}
\si E_A(\si)^{\frac{-1+it}{2}}E_A(\rho)^{\frac{1-it}{2}}
\Big)
\end{align*}
Under the assumption of Theorem \ref{pssa}, $E_A(\si)=\si$ and $E_A^\dagger(\B)\subset \R$,
\begin{align*}
c(t)&=\sup_{b\in B} \tau_M(b
E_A(\rho))=\sup_{b\in B} \tau_M(E_A^\dagger(b)
\rho)=\sup_{b\in B} \tau_M\Big(E_R^\dagger(b)
\rho
\Big)
= 1
\end{align*}
This recovers the assertion of Theorem \ref{pssa}
}
\end{exam}


\bibliography{UCR}

\bibliographystyle{plain}
\end{document}